\documentclass[aps, prb, twocolumn, superscriptaddress, amsmath,  tightenlines, longbibliography]{revtex4-1}

\usepackage{dcolumn}
\usepackage{graphicx}
\usepackage{mathrsfs}
\usepackage{subfigure}
\usepackage{booktabs}
\usepackage{amsmath}
\usepackage{physics}
\usepackage{dsfont}
\usepackage{amstext}
\usepackage{amssymb}
\usepackage{amsbsy}
\usepackage{bbm}
\usepackage{amsthm}
\usepackage{graphicx}
\usepackage{color}
\usepackage[colorlinks,citecolor=blue]{hyperref}

\usepackage{url}
\usepackage[colorlinks]{hyperref}
\hypersetup{%
	plainpages=true,
	breaklinks=true,       
	hypertexnames=false,  
	pageanchor=true,
	colorlinks=true,
	linkcolor={blue},
	citecolor={red},
	urlcolor={blue},
	anchorcolor={black}
}

 \makeatletter

\newcommand{\Rmnum}[1]{\expandafter\@slowromancap\romannumeral #1@}
\makeatother

\begin{document}
	
\title{Interaction-Induced Higher-Order Topological Insulator via Floquet Engineering}
\author{Chun-Ping Su}
\thanks{These authors contributed equally}
\affiliation{School of Physics and Optoelectronics, South China University of Technology,  Guangzhou 510640, China}
\author{Zhao-Fan Cai}
\thanks{These authors contributed equally}
\affiliation{School of Physics and Optoelectronics, South China University of Technology,  Guangzhou 510640, China}
\author{Tao Liu}
\email[E-mail: ]{liutao0716@scut.edu.cn}
\affiliation{School of Physics and Optoelectronics, South China University of Technology,  Guangzhou 510640, China}

\date{{\small \today}}


\begin{abstract}
	Higher-order topological insulators have attracted significant interest in both static single-particle and many-body lattice systems. While periodically driven (Floquet) higher-order topological phases have been explored at the single-particle level, the role of interactions in such systems remains less understood. In this paper, we extend previous studies by investigating interaction-induced higher-order topological phases through Floquet engineering.	To achieve this, we construct an extended Bose-Hubbard model on a square lattice subjected to periodic driving. We demonstrate the emergence of interaction-induced normal Floquet second-order topological corner states for doublons (i.e., bound boson pairs) from a trivial phase, which exhibit robustness against disorder.  Notably, beyond the normal phase, we reveal an interaction-induced anomalous Floquet second-order topological phase, where in-gap corner states of doublons emerge within the $\pi/T$ gap ($T$ being the driving period).	Our model, accessible with state-of-the-art ultracold atom techniques, provides a platform for realizing interaction-driven higher-order topological phases uniquely enabled by periodic driving, with no direct counterparts in static or single-particle systems.
\end{abstract}

\maketitle
	
\section{Introduction}\label{section1}

In recent years, there has been significant interest in exploring higher-order topological phases \cite{PhysRevLett.110.046404, Benalcazar61, PhysRevB.96.245115, PhysRevLett.119.246401,PhysRevLett.119.246402}. These novel topological states go beyond the conventional bulk-boundary correspondence. Instead, a  $d$-dimensional $n$th-order  topological phase hosts gapless states on its $(d-n)$-dimensional boundary, enriching the landscape of topological phases \cite{Xie2021}. Higher-order topological phases of matter have been extensively studied across various branches of topological physics within static single-particle  systems \cite{ Peterson2018, Serra-Garcia2018,   arXiv:1801.10053,  PhysRevLett.120.026801,  TitusSciAdv2018, Ni2018, Xue2018, arXiv:1801.10050, arXiv:1802.02585,   PhysRevLett.123.216803,  Mittal2019, ElHassan2019,  Zhang2019, PhysRevResearch.2.033029,  PhysRevB.101.241104,PhysRevLett.124.036803, PhysRevLett.124.063901, PRBXYZhu2018, arXiv:1803.08545,    arXiv:1806.07002, PhysRevLett.121.196801, PhysRevLett.123.177001,  PhysRevLett.122.236401, PhysRevB.99.020508, PhysRevLett.122.076801,  PhysRevLett.122.187001, PhysRevLett.123.156801,PhysRevB.103.085408,PhysRevB.103.L201115, PhysRevLett.125.146401,PhysRevB.107.125118,PhysRevB.107.165403, PhysRevLett.124.063901,PhysRevLett.124.036803,Kang2023,Yang2024,PhysRevLett.126.206404,PhysRevResearch.2.012067,PhysRevB.105.L201301}.

Higher-order topological phases can extend beyond static systems, with time serving as an additional dimension that gives rise to fascinating band structure features \cite{Rudner2020}. In a Floquet system, the Hamiltonian $ \mathcal{H}(t)$  preserves time periodicity, i.e., $ \mathcal{H}(t) = \mathcal{H}(t+T) $, due to the periodic driving in period $T$. This periodic driving  can turn a trivial band insulator into a topological insulator \cite{Rudner2020,PhysRevLett.112.026805, Lindner2011, Maczewsky2020,Pyrialakos2022}. Moreover, going beyond their static counterparts,  the Floquet systems can host anomalous topological phases due to the nontrivial winding of the quasienergy spectrum \cite{rudner2013anomalous, Peng2016, maczewsky2017observation, PhysRevB.96.195303, PhysRevB.99.235112, Wintersperger2020, Wintersperger2020}. A typical example is the emergence of chiral edge states in anomalous Floquet topological insulators, where the bulk bands have a vanishing Chern number, thereby violating the conventional bulk-edge correspondence \cite{Rudner2020}. In the realm  of higher-order topology, Floquet higher-order topological insulators have recently garnered significant attention and investigation \cite{PhysRevB.99.195426, PhysRevLett.123.016806,  hu2020dynamical, huang2020floquet, Yu2021, PhysRevB.103.115308, PhysRevB.101.235403, PhysRevResearch.2.043431, zhu2022time, PhysRevB.110.125427}.

The interplay between topology and many-body interactions gives rise to a rich variety of novel phases beyond conventional single-particle topological states. In first-order topological systems, interactions can drive exotic phenomena such as interaction-driven topological insulators from trivial phases, fractional Chern insulators, and topological Mott insulators \cite{Hohenadler2013, WitczakKrempa2014, PhysRevLett.114.177202}. More recently, extensive efforts have been dedicated to understanding how many-body interactions influence higher-order topological phases \cite{PhysRevB.98.235102,PhysRevB.99.235132, PhysRevB.101.085137, PhysRevB.102.045110, PhysRevB.106.L241115, PhysRevB.106.155116, PhysRevLett.123.196402, PhysRevLett.127.176601, PhysRevResearch.2.012009, PhysRevB.102.041126, PhysRevLett.131.263001}.

\begin{figure}[!tb]
	\centering
	\includegraphics[width=8.7cm]{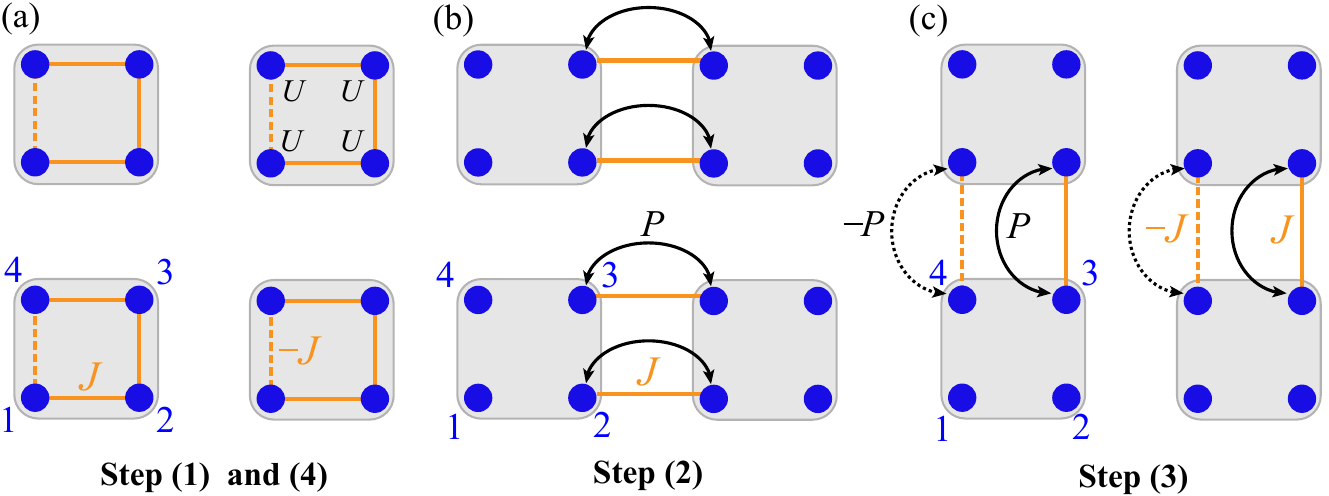}
	\caption{Schematic of an interaction-induced HOTI on a square lattice using a four-step driving scheme over a period $T$: (a) $\mathcal{H}_0$ governs steps 1 and 4, (b) $\mathcal{H}_x$ governs step 2, and (c) $\mathcal{H}_y$ governs step 3. Each unit cell (gray boxes) contains four sublattice sites labeled $\alpha \in {1, 2, 3, 4}$. During each step, boson-boson interactions occur with onsite interaction strength $U$, acting over a time interval of $T/4$. Hopping between sites is characterized by two parameters: $J$ for both inter-cell and intra-cell hopping (orange lines) and $P$ for inter-cell two-boson hopping (curved double arrows). The solid and dashed lines distinguish positive and negative signs for the hopping terms, respectively.}\label{fig1}
\end{figure}
\begin{figure*}[!tb]
	\centering
	\includegraphics[width=18cm]{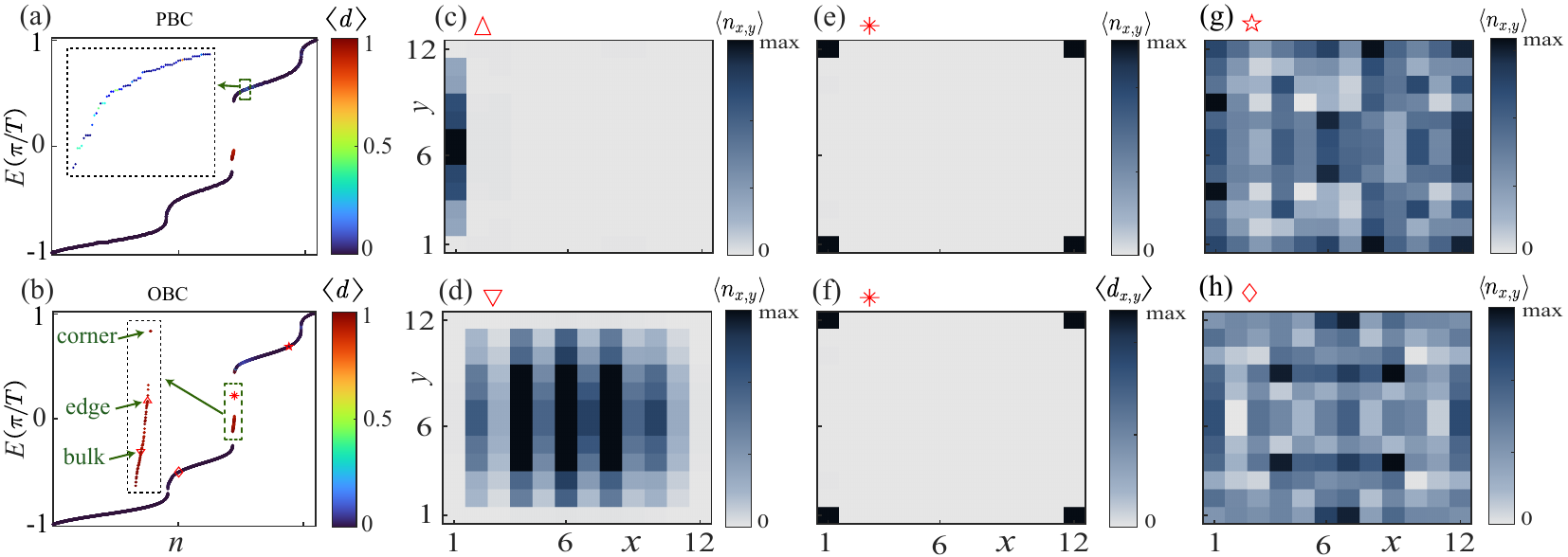}
	\caption{(a,b) Quasienergies spectra $E$ of the Floquet Hamiltonian $\mathcal{H}_\textrm{F}$ under (a) PBCs and (b) OBCs along both $x$ and $y$ directions.  The color scale represents the total double-occupation bosonic density $\langle d \rangle = \sum_{x, y} d_{x,y}  $.  The lower band of the PBC spectrum corresponds to single occupation, the middle band to double occupation, and the upper band consists of states with a mix of single and double occupations, as highlighted in the magnified view in (a).  Panels (c–h) display the single-occupation and double-occupation density distributions, $\langle n_{x,y} \rangle$ and $\langle d_{x,y} \rangle$, for specific quasienergies highlighted in (b): (c) edge states of doublons, (d) bulk states of doublons, (e, f) four-corner states of doublons, and (g, h) bulk states of scattering states. The parameters used here are $J=1$, $P = 3$, $U  = 8$, and $T=1$ with lattice size $L\times L=12\times12$.}\label{fig2}
\end{figure*}

While fully many-body systems present significant theoretical challenges, valuable insights can be gained by studying simpler few-body systems. A particularly interesting scenario involves two interacting particles in a lattice, which can form a bound state known as a doublon \cite{winkler2006repulsively}. Such bound states arise for both attractive and repulsive interactions and effectively behave as composite particles \cite{winkler2006repulsively}. The interplay between topology and two-body interactions has recently attracted growing interest, revealing novel interaction-induced first-order topological states \cite{salerno2018topological, lyubarov2019edge, lin2020interaction, stepanenko2020interaction, gorlach2017topological, olekhno2020topological, stepanenko2020topological}. Most strikingly, higher-order topological insulators arising from two interacting bosons have also been investigated in static systems \cite{PhysRevLett.128.213903}. Given these developments, a natural question arises: How does the interplay between two-body interactions and periodic driving determine second-order topological phases? Can anomalous Floquet second-order topological corner states emerge in such two-body systems?

In this work, we take an important step in exploring the interplay between periodic driving and many-body interactions in two-dimensional (2D) lattices by constructing a time-periodic extended Bose-Hubbard model on a square lattice. We focus on higher-order topological phases in the two-body interaction regime, where the system remains topologically trivial at the single-boson level. Our results reveal that the Floquet Hamiltonian hosts normal in-gap second-order topological corner states of doublons (i.e., bound boson pairs). In the high-frequency approximation and strong-interaction limit, we derive an effective doublon Hamiltonian, showing that the periodically driven interacting lattice maps onto a generalized Benalcazar-Bernevig-Hughes (BBH) model \cite{Benalcazar61}, confirming its topological nature. Moreover, we identify an anomalous Floquet second-order topological phase, where in-gap doublon corner states emerge within the $\pi/T$ gap ($T$ being the driving period). Importantly, both normal and anomalous Floquet second-order topological corner states are shown to be robust against disorder, demonstrating the stability of these interaction-induced phases enabled by Floquet engineering.

The structure of this paper is as follows. In Sec.~\ref{section2}, we introduce a two-dimensional square lattice model subject to periodic driving, incorporating onsite many-body interactions and two-boson pairing hopping. In Sec.~\ref{section3}, we investigate the emergence of interaction-induced normal Floquet second-order topological corner states. To elucidate the underlying mechanism, we derive an effective doublon Hamiltonian within the high-frequency approximation and strong-interaction limit. Additionally, we examine the influence of Tamm-Shockley states and disorder. We further explore the anomalous Floquet second-order topological states that emerge within the $\pi/T$ gap. Finally, in Sec.~\ref{section4}, we summarize our findings and present our conclusions.

\section{Model }\label{section2}

We study an interacting tight-binding model on a square lattice with four sublattices, driven periodically in time. The driving follows a four-step sequence, as illustrated in Fig.~\ref{fig1}. Over a single period $T$, the time-dependent Hamiltonian $\mathcal{H}(t)$ is expressed as
\begin{align}\label{Ht}
	\mathcal{H}(t) = \begin{cases}
		&\mathcal{H}_0,~~~t\in T_1,~T_4, \\
		&\mathcal{H}_x,~~~t\in T_2, \\
		&\mathcal{H}_y,~~~t\in T_3,
	\end{cases}
\end{align}
where the time interval $T_{s}=[(s-1)T/4, ~sT/4]$ ($s=1,2,3,4$), and the Hamiltonian in each step is written as
\begin{align}\label{H0}
	\mathcal{H}_{0}  = & ~ J \sum_{x, y}\left(a^\dagger_{x, y, 1} a_{x,y,2} + a^\dagger_{x, y,3} a_{x,y,4} + \textrm{H.c.}\right) \nonumber \notag \\
	&+J \sum_{x, y}\left(a^\dagger_{x, y,2} a_{x, y,3} - a^\dagger_{x, y,4} a_{x, y, 1} + \textrm{H.c.}\right) \nonumber \notag \\
	& + U \sum_{x, y}\sum_{\alpha \in\{1,2,3,4\}}  n_{x, y, \alpha} n_{x, y, \alpha},
\end{align}
\begin{align}\label{Hx}
	\mathcal{H}_{x} = &~  J \sum_{x,y} \left(a^\dagger_{x+1,y,1} a_{x, y,2} + a^\dagger_{x+1,y,4} a_{x, y,3}+ \textrm{H.c.} \right)   \nonumber \notag \\
	&+ \frac{P}{2} \sum_{x, y}\left(a^\dagger_{x+1,y,1} a^\dagger_{x+1,y,1} a_{x, y,2} a_{x,y,2} + \textrm{H.c.}\right) \nonumber  \notag \\ 
	&+ \frac{P}{2} \sum_{x, y}\left(a^\dagger_{x+1,y, 4} a^\dagger_{x+1,y,4} a_{x,y,3} a_{x, y,3} + \textrm{H.c.}\right) \nonumber  \notag \\ 
	&+ U \sum_{x, y}\sum_{\alpha \in\{1,2,3,4\}}  n_{x, y, \alpha} n_{x, y, \alpha},
\end{align}
and
\begin{align}\label{Hy}
	\mathcal{H}_{y} =&~  J \sum_{x, y} \left(- a^\dagger_{x,y+1,1} a_{x, y,4} + a^\dagger_{x,y+1,2} a_{x, y,3}+ \textrm{H.c.} \right) \nonumber 
	\notag \\ 
	&- \frac{P}{2} \sum_{x, y}\left(a^\dagger_{x,y+1,1} a^\dagger_{x,y+1,1} a_{x, y,4} a_{x,y,4} + \textrm{H.c.}\right) \nonumber \notag \\
	&+ \frac{P}{2} \sum_{x, y}\left(a^\dagger_{x,y+1,2} a^\dagger_{x,y+1,2} a_{x, y,3} a_{x, y,3} + \textrm{H.c.}\right) \nonumber \notag \\
	&+ U \sum_{x, y}\sum_{\alpha \in\{1,2,3,4\}}  n_{x, y, \alpha} n_{x, y, \alpha}.
\end{align} 
Here, $a^\dagger_{x, y, \alpha}$ represents the creation operator for a boson on the sublattice  $\alpha\in \{1,2,3,4\}$ at site $(x,y)$ within a unit cell.  The intra-cell and inter-cell hopping amplitudes are both of equal strength, denoted by $J$. The parameter $U$ characterizes the onsite boson-boson interaction strength, while $P$ describes the strength of inter-cell two-boson hopping. The Hamiltonian $\mathcal{H}(t)$ in Eq.~(\ref{Ht}) is second-order topologically trivial in the single-particle regime, where $U=P=0$.

\begin{figure*}[!tb]
	\centering
	\includegraphics[width=18cm]{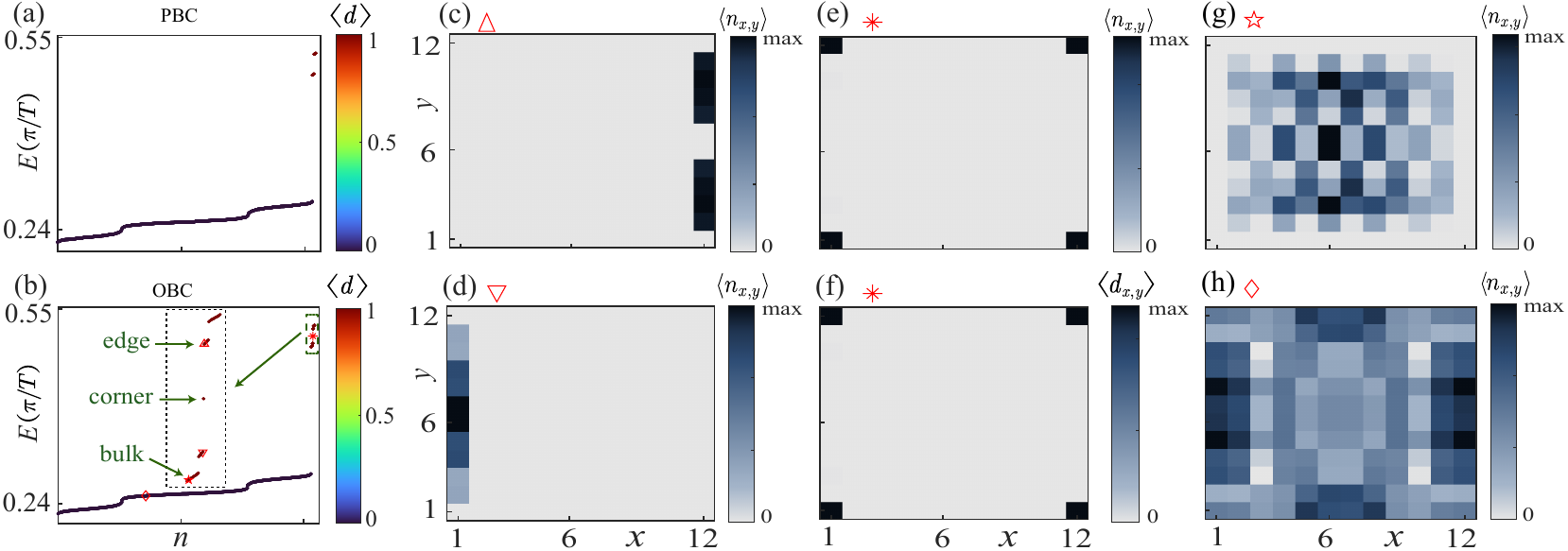}
	\caption{(a,b) Quasienergies spectra $E$ of the Floquet Hamiltonian $\mathcal{H}_\textrm{F}$ under (a) PBCs and (b) OBCs along both $x$ and $y$ directions.  The color scale represents the total double-occupation bosonic density $\langle d \rangle = \sum_{x, y} d_{x,y}  $. Panels (c–h) display the single-occupation and double-occupation density distributions, $\langle n_{x,y} \rangle$ and $\langle d_{x,y} \rangle$, for specific quasienergies highlighted in (b): (c,d) edge states of doublons, (e, f) four-corner states of doublons, (g) bulk states of doublons, and (h) bulk states of scattering states. The parameters used here are $J=1$, $P = 3$, $U  = 8$, and $T=0.05$ with lattice size $L\times L=12\times12$.}\label{fig3}
\end{figure*}

According to the  Floquet theorem \cite{Rudner2020}, a time-periodic Hamiltonian $\mathcal{H}(t) = \mathcal{H}(t+T)$, with the driving period $T = 2 \pi /\omega$, is governed  by  Schr$\ddot{\text{o}}$dinger equation $i\partial_t |\psi_m(t)\rangle  =  \mathcal{H}(t) |\psi_m(t)\rangle$. The system admits a complete set of orthogonal solutions of the form $|\psi_m(t)\rangle = e^{-i E_m t} |u_m(t)\rangle$,   where $|u_m(t)\rangle$ is periodic in time, satisfying $|u_m(t)\rangle = |u_m(t+T)\rangle$, and  $E_n$ denotes the quasienergy. In this work, we focus on the stroboscopic dynamics described by the time-independent effective Floquet Hamiltonian $\mathcal{H}_\textrm{F}$, defined as
\begin{align}\label{UnitaryEvolution}
U_\text{F} \equiv 	U(T)   =     \mathcal{T} e^{-i\int_{0}^{T} \mathcal{H}(t') dt'} = e^{-i\mathcal{H}_\textrm{F} T } ,
\end{align}
where $\mathcal{T}$ is the time-ordering operator, and $U_\text{F}  |\psi_m(0)\rangle = e^{-i E_m T} |\psi_m(0)\rangle $.  

Furthermore, according to Eqs.~(\ref{Ht}-\ref{Hy}), the   stroboscopic dynamics can be rewritten as
\begin{align}\label{UF2}
	 U_\text{F} =e^{-i \mathcal{H}_{0}\frac{T}{4}} e^{-i \mathcal{H}_{y}\frac{T}{4}} e^{-i \mathcal{H}_{x}\frac{T}{4}} e^{-i \mathcal{H}_{0}\frac{T}{4}}.
\end{align}

\section{Results}\label{section3}

\subsection{Normal Floquet topological corner states}

In the absence of interactions, the time-independent effective Floquet Hamiltonian $\mathcal{H}_\textrm{F}$ does not exhibit higher-order topological phases. We now turn our attention to the topological characteristics of the quasienergy spectrum under the influence of interactions. Since the Hamiltonian $\mathcal{H}(t) $ preserves   $U(1)$ symmetry, we focus on studying the topological phases within the double-excitation subspace. In this context, we numerically calculate the quasienergy spectrum $E$, as well as the single-occupation and double-occupation particle densities, defined as 
\begin{align}\label{particledensityn}
	\langle n_{x,y} \rangle = \langle \psi_m | n_{x,y} |\psi_m\rangle,
\end{align}
and
\begin{align}\label{particledensityndoublon}
	\langle d_{x,y} \rangle = \langle \psi_m | \frac{1}{2}a_{x,y}^\dagger a_{x,y}^\dagger a_{x,y} a_{x,y} |\psi_m\rangle,
\end{align}
where $\ket{\psi_m} (m=1,2,3,\cdots)$ is the $m$th normalized eigenvector of the Floquet Hamiltonian $\mathcal{H}_\textrm{F}$.

Figure \ref{fig2}(a,b)  illustrates the quasienergy spectra $E$ of the Floquet Hamiltonian under periodic boundary conditions (PBCs) and open boundary conditions (OBCs) along both $x$ and $y$ directions for $T=1$.  The color scale indicates the total double-occupation bosonic density $\langle d \rangle = \sum_{x, y} d_{x,y}  $, where $\langle d \rangle = 0$ corresponds to states where two bosons occupy different sites, while $\langle d \rangle = 1$ represents states where both bosons occupy the same site. Under PBCs, the eigenspectrum is distinctly divided into three main regions: a lower-energy scattering band with $\langle d \rangle = 0$, a middle doublon band characterized by $\langle d \rangle = 1$,  and an upper band containing states with mixed single and double occupations. The scattering states correspond to superpositions of two-particle configurations with particles localized on different lattice sites. In contrast, the doublon bands consist of bound bosonic states, where both bosons are co-localized on the same site. 

When the boundaries are open along both the $x$ and $y$ directions for $T=1$, two types of in-gap states emerge [see Fig.~\ref{fig2}(b)]. The first type is the doublon edge states, which are localized along the one-dimensional edges [see Fig.~\ref{fig2}(b,c)]. The edge modes touch the doublon bulk bands, with the extended state distribution of the bulk bands shown in Fig.~\ref{fig2}(d). The second type is four doublon corner states, which are localized at the zero-dimensional corners [see Fig.~\ref{fig2}(b,e,f)]. The doublon nature of the corner states can be further confirmed by comparing the single-occupation and double-occupation density distributions, as shown in Fig.~\ref{fig2}(e,f).  The single-occupation distribution,  $\langle n_{x,y} \rangle$, reflects the probability of finding a boson at site $(x,y)$, regardless of whether the second boson is at the same site or elsewhere. In contrast, the double-occupation distribution, $\langle d_{x,y} \rangle$, specifically measures the probability that both bosons occupy the same site. If these two distributions are identical, it implies that the bosons are always bound together at the same site, confirming the formation of doublons. In addition, the lower-energy scattering bands and higher-energy bands are extended, with their state distributions shown in Fig.~\ref{fig2}(g,h).

For a moderate driving frequency of $\omega=2\pi/T = 2\pi$ [e.g., see Fig.~\ref{fig2}],  the corner modes appear as in-gap states located between the doublon bulk band and the higher-energy mixed bulk band. When the driving frequency is further increased, such as $\omega=2\pi/T = 2\pi/0.05$ in Fig.~\ref{fig3},  the corner modes emerge as in-gap states situated between two doublon bulk bands.  As shown in Fig.~\ref{fig3}(a), under the PBCs, the eigenspectrum is distinctly divided into three main regions: a lower-energy scattering band characterized by $\langle d \rangle = 0$, and two doublon bands (middle and higher-energy) with $\langle d \rangle = 1$. When the OBCs  are applied along both the $x$ and $y$ directions, the spectrum reveals two edge bands and four corner states positioned between the two doublon bulk bands. The edge states are localized along the one-dimensional edges [see Fig.~\ref{fig3}(b,c,d)], while the corner states are confined to the zero-dimensional corners [see Fig.~\ref{fig3}(b,e,f)]. In contrast, the doublon and scattering bulk bands remain extended, as illustrated in Fig.~\ref{fig3}(g,h), respectively.

\begin{figure}[!tb]
	\centering
	\includegraphics[width=9cm]{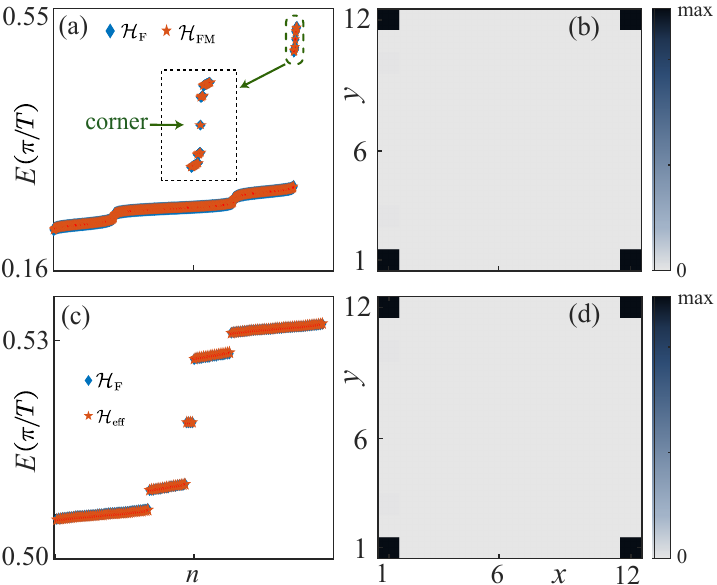}
	\caption{(a) Quasienergy $E$ of the Floquet Hamiltonian $\mathcal{H}_\textrm{F}$ (blue diamonds) and the  effective Hamiltonian $\mathcal{H}_\text{FM}$ (orange stars) in the high-frequency approximation under OBCs in both the $x$ and $y$ directions, and (b) the corresponding state distributions of the corner modes using $\mathcal{H}_\text{FM}$. (c) Quasienergy $E$ of the Floquet Hamiltonian $\mathcal{H}_\textrm{F}$ (blue diamonds) and the  effective  Hamiltonian $\mathcal{H}_\textrm{eff}$ (orange stars) of doublons in the high-frequency approximation and strong-interaction limit   under OBCs in both the $x$ and $y$ directions, and (d) the corresponding state distributions of the corner modes   using $\mathcal{H}_\textrm{eff}$.  The parameters used  are $J=1$, $P = 3$, $U  = 8$, and $T=0.05$ with lattice size $L\times L=12\times12$.}\label{fig4}
\end{figure}

\subsection{Topological nature of normal Floquet corner states in high-frequency driving}\label{section5}

To elucidate the mechanism of interaction-induced topological second-order corner states through Floquet engineering in the high driving-frequency case, we derive the effective time-independent Hamiltonian for doublons by employing a high-frequency approximation and strong-interaction limit. When the driving frequency $\omega = 2\pi/T$ significantly exceeds all other characteristic energy scales of the system, e.g., see Fig.~\ref{fig3}, the   effective time-independent Hamiltonian $\mathcal{H}_\textrm{FM}$ can be obtained using the Floquet-Magnus expansion \cite{Bukov2015, Eckardt2015, PhysRevB.93.144307,PhysRevB.98.104303,PhysRevLett.121.080401,RevModPhys.89.011004,PhysRevResearch.6.L042033}. This approach captures the key features of the Floquet dynamics and enables a clear understanding of the emergent corner states in the high driving-frequency case. Furthermore, when the onsite interaction strength $U$ is much larger than the hopping strength, we can further utilize the quasi-degenerate perturbation theory \cite{Bir1974,CCohenTannoudji1Atom},

In the high-frequency approximation, the time-independent effective Hamiltonian $\mathcal{H}_\text{FM}$ can be obtained using the Floquet-Magnus expansion \cite{Bukov2015, Eckardt2015, PhysRevB.93.144307, PhysRevB.98.104303, PhysRevLett.121.080401, RevModPhys.89.011004, PhysRevResearch.6.L042033}. In the zeroth-order approximation, it is given by
\begin{align}\label{relationsHFM}
	\mathcal{H}_\text{FM} = \frac{1}{T}\int_{0}^{T} \mathcal{H}(t) \,dt=(2\mathcal{H}_{0}+\mathcal{H}_{x}+\mathcal{H}_{y})/4.
\end{align}
\begin{figure*}[!tb]
	\centering
	\includegraphics[width=18cm]{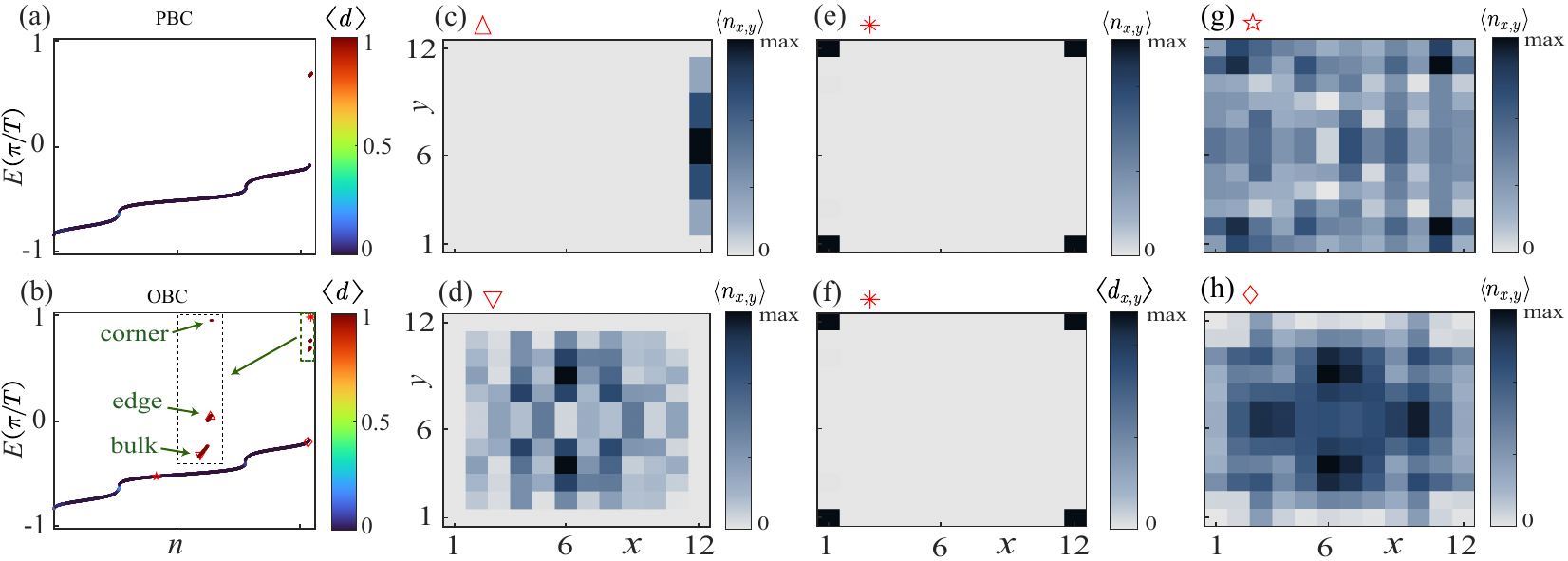}
	\caption{(a,b) Quasienergies spectra $E$ of the Floquet Hamiltonian $\mathcal{H}_\textrm{F}$ under (a) PBCs and (b) OBCs along both $x$ and $y$ directions.  The color scale represents the total double-occupation bosonic density $\langle d \rangle = \sum_{x, y} d_{x,y}  $. Panels (c–h) display the single-occupation and double-occupation density distributions, $\langle n_{x,y} \rangle$ and $\langle d_{x,y} \rangle$, for specific quasienergies highlighted in (b): (c) edge states of doublons, (d) bulk states of doublons,  (e,f) four-corner states of doublons, (g,h) bulk states of scattering states. The parameters used here are $J=0.5$, $P = 3$, $U  = 2.35$, and $T=1$ with lattice size $L\times L=12\times12$.}\label{fig5}
\end{figure*}

Figure \ref{fig4}(a)  presents the quasienergy spectra $E$ of the Floquet Hamiltonian $\mathcal{H}_\textrm{F}$ (blue diamonds) and the  effective Hamiltonian $\mathcal{H}_\text{FM}$ (orange stars) in the high-frequency approximation under OBCs in both the $x$ and $y$ directions. Additionally, it depicts the corresponding state distributions of the corner modes obtained using  $\mathcal{H}_\text{FM}$ in Fig.~\ref{fig4}(b). The results demonstrate that the effective Hamiltonian $\mathcal{H}_\text{FM}$ closely matches the original Floquet Hamiltonian $\mathcal{H}_\textrm{F}$ in the high driving frequency.
 
Furthermore, when the driving frequency $\omega = 2\pi/T$ is large, and the onsite interaction strength satisfies $\abs{U} \gg \abs{P}, J$, the doublon bands become well-separated from the scattering bands, as illustrated in Fig.~\ref{fig3}. In this strong-interaction limit, we further derive the effective Hamiltonian $\mathcal{H}_\textrm{eff}$ of doublons by utilizing quasi-degenerate second-order perturbation theory \cite{Bir1974,CCohenTannoudji1Atom}. As illustrated in Appendix \ref{AppendixA},  we define a doublon subspace spanned by $d^\dagger_{x, y, \alpha} |0\rangle$, with $d^\dagger _{x, y, \alpha} \equiv a^\dagger_{x, y, \alpha} a^\dagger_{x, y, \alpha}/ \sqrt2 $,  the effective doublon Hamiltonian $\mathcal{H}_\text{eff}$ can be written as 
\begin{align}\label{H30}
	\mathcal{H}_\text{eff}   =  & \frac{J^2}{4U} \sum_{x, y}\left(d^\dagger_{x, y, 1} d_{x,y,2} + d^\dagger_{x, y,3} d_{x,y,4} + \textrm{H.c.}\right) \nonumber \\
	&+ \frac{J^2}{4U} \sum_{x, y}\left(d^\dagger_{x, y,2} d_{x, y,3} + d^\dagger_{x, y,4} d_{x, y, 1} + \textrm{H.c.}\right) \nonumber \\
	& + \left(\frac{J^2}{16U}+\frac{P}{4}\right) \sum_{x,y} \left(d^\dagger_{x+1,y,1} d_{x, y,2}  + \textrm{H.c.} \right)       \nonumber \\ 
	&  + \left(\frac{J^2}{16U}+\frac{P}{4}\right) \sum_{x,y} \left(d^\dagger_{x+1,y,4} d_{x, y,3} + \textrm{H.c.} \right)       \nonumber \\ 
	& +  \left(\frac{J^2}{16U}+\frac{P}{4}\right) \sum_{x,y} \left( d^\dagger_{x,y+1,2} d_{x, y,3}+ \textrm{H.c.} \right)       \nonumber \\ 
	& + \left(\frac{J^2}{16U}-\frac{P}{4}\right) \sum_{x,y} \left( d^\dagger_{x,y+1,1} d_{x, y,4} + \textrm{H.c.} \right)       \nonumber \\
	& +  \sum_{x, y}\sum_{\alpha \in\{1,2,3,4\}} U_\text{eff}(x,y) d^\dagger_{x, y, \alpha} d_{x,y,\alpha}.            
\end{align}
where $U_\text{eff}(x,y) = 4U + 5J^2/(8U)$ under the PBCs.  Under the OBCs, $U_\text{eff}(x,y)$ depends on whether $(x,y)$ is in the bulk, on the edge, or at the corner under the OBCs, and can be expressed as
\begin{align}\label{Ueff1}
	U_\text{eff}(x,y) = \begin{cases}
		4U + \frac{5J^2}{8U}, &~\text{if}~ (x,y) ~\text{in the bulk}, \\
		4U + \frac{9J^2}{16U}, &~\text{if}~ (x,y) ~\text{on the edge}, \\
		4U + \frac{J^2}{2U}, &~\text{if}~ (x,y) ~\text{at the corner}.
	\end{cases} 
\end{align}

Figure \ref{fig4}(c,d) shows quasienergies spectra $E$ of the Floquet Hamiltonian $\mathcal{H}_\textrm{F}$ (blue diamonds) and the  effective  Hamiltonian $\mathcal{H}_\textrm{eff}$ (orange stars) of doublons in the high-frequency approximation and strong-interaction limit   under OBCs in both the $x$ and $y$ directions, and the corresponding state distributions of the corner modes  using $\mathcal{H}_\textrm{eff}$. The results indicate that the time-independent effective single-particle Hamiltonian 
$\mathcal{H}_\textrm{eff}$ effectively captures the doublon eigenspectrum of the Floquet Hamiltonian $\mathcal{H}_\textrm{F}$ in the high-frequency approximation and strong-interaction limit.  Thus, we can analyze the topological origin of the interaction-driven topological corner states through Floquet engineering based on $\mathcal{H}_\textrm{eff}$ in the high driving-frequency case.

According to Eq.~(\ref{H30}),   the effective Hamiltonian $\mathcal{H}_\text{eff}$ describes a doublon as a quasiparticle on a square lattice. This system realizes a generalized Benalcazar-Bernevig-Hughes (BBH) model \cite{Benalcazar61, PhysRevB.96.245115}, featuring dimerized hopping  along both the $x$ and $y$ directions,  akin to a two-dimensional Su-Schrieffer-Heeger (SSH) model. Moreover, the last second term introduces a $\pi$-flux in alternating plaquettes due to its negative sign. Consequently, the effective Hamiltonian $\mathcal{H}_\text{eff}$ hosts topological second-order corner states, reflecting the topological nature of normal Floquet corner states under high-frequency driving.  In addition, as shown in Eqs.~(\ref{H30}) and (\ref{Ueff1}), under OBCs, the difference in onsite potentials between boundary and bulk sites can lead to two-particle Tamm-Shockley states \cite{Shockley1932,PhysRev.56.317}, which are corner-localized but lack a topological origin. However, as demonstrated in Appendix \ref{AppendixB}, the in-gap states localized at the corners, discussed here, do not belong to this category. Instead, they emerge from the topological nature of the bulk bands.

\subsection{Anomalous Floquet topological corner states}

Floquet topological insulators can host normal boundary states associated with bulk states winding around crystal momentum space, whose topological properties are described by an effective Hamiltonian. However, unlike static systems, Floquet systems also support anomalous topological phases, which are characterized by edge states at the quasienergy gap of $\pi/T$ \cite{Rudner2020}. These anomalous phases challenge the conventional understanding of bulk-boundary correspondence, as they can exhibit robust edge states even when the bulk bands have trivial topological invariants, such as a vanishing Chern number \cite{Rudner2020}. The difficulty in fully capturing the topology of a Floquet-Bloch system using an effective Hamiltonian stems from the unique nature of quasienergy. In a static lattice,  energy spectra are bounded, meaning edge modes cannot extend beyond the lowest or highest bulk bands \cite{Rudner2020}. In contrast, the Floquet spectrum is not bounded but instead periodic and confined within a compact Floquet-Brillouin zone, typically defined in the interval $[-\pi/T,~\pi/T]$\cite{rudner2013anomalous}. This periodicity allows quasienergy bands to wind around the Floquet-Brillouin zone an integer number of times as the momentum traverses the Brillouin zone. Such winding is a hallmark of Floquet systems and has no direct analog in static systems, where energy is defined on an open domain.

In the presence of interactions, periodic driving can also induce anomalous boundary states at the quasienergy gap of $\pi/T$. Figure \ref{fig5}(a,b) shows the quasienergy spectra $E$ of the Floquet Hamiltonian under periodic boundary conditions (PBCs) and open boundary conditions (OBCs) along both $x$ and $y$ directions for $T=1$ and $J=0.5$. In contrast to quasienergy spectra in Fig.~\ref{fig2}, the band gap at  $\pi/T$ is open [see Fig.~\ref{fig5}(a)] under PBCs. When the boundaries are open along both the $x$ and $y$ directions, anomalous in-gap states of doublons emerge [see Fig.~\ref{fig5}(b)]. These in-gap states are edge states [see Fig.~\ref{fig5}(c)], and second-order topological corner states [see Fig.~\ref{fig5}(e,f)],  which appear above the bulk states [see Fig.~\ref{fig5}(d)]. The lower-energy scattering bands  are extended, with their state distributions shown in Fig.~\ref{fig5}(g,h). These results indicate the emergence of interaction-induced anomalous Floquet second-order topological states in our model.

\begin{figure}[!tb]
	\centering
	\includegraphics[width=8.7cm]{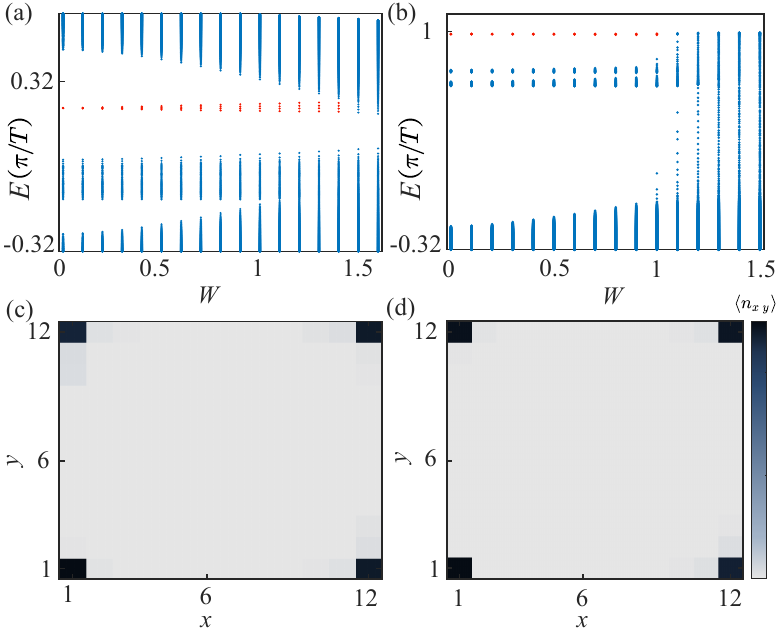}
	\caption{Quasienergies $E$ of the Floquet Hamiltonian under OBCs along both $x$ and $y$ directions, subject to random disorder applied to the single-particle hopping terms, as a function of the disorder strength $W$, where only part of quasienergies is shown for each $W$. The red dots indicate the in-gap corner states. The parameters are (a) $J = 1$, $P = 3$, $U  = 8$, $T=1$, and (b) $J=0.5$, $P = 3$, $U  = 2.35$, $T=1$. The panel  (c) shows the corresponding four corner states marked by red dots for $W=1.2$ in (a). The panel  (d) shows the corresponding four corner states marked by red dots for $W=1$ in (b). The averages are calculated over 400 disorder realizations.}\label{fig6}
\end{figure}

\subsection{Roust against Disorder}

The periodically driven interacting model studied here supports both normal and anomalous Floquet second-order corner states. To examine their robustness against local disorder due to topological protection, we introduce random disorder in the single-particle hopping terms. Specifically, the single-particle hopping amplitude $J$ in the four-step driving process is modified as $J\rightarrow J + V_\text{dis}$ in Eqs.~(\ref{Ht})-(\ref{Hy}). Here, the random disorder $V_\text{dis}$ is uniformly sampled from the range $[-W/2,W/2]$, with $W$ being disorder strength.

Figure \ref{fig6} shows  quasienergies   $E$ of the Floquet Hamiltonian under OBCs along both $x$ and $y$ directions, subject to random disorder applied to the single-particle hopping terms, as a function of the disorder strength $W$. For interaction-induced normal Floquet second-order topological corner states, the in-gap corner states persist until a much stronger disorder is reached [red dots in Fig. \ref{fig6}(a)], at which point they merge into the bulk states. As the disorder strength increases, the fourfold degeneracy of the corner states is lifted. However, these in-gap states remain well localized at the corners [see Fig. \ref{fig6}(c)]. Moreover, for anomalous Floquet second-order topological states, the corner states remain also well within the band gap at $\pi/T$, and retain their degeneracy even in the presence of disorder [see Fig. \ref{fig6}(b)]. Furthermore, these in-gap states remain localized at the corners even at large disorder strengths [see Fig. \ref{fig6}(d)]. However, at sufficiently strong disorder, the gap at $\pi/T$ eventually closes, marking the breakdown of the  gapped topological phase. Notably, in the single-particle case, disorder can give rise to a gapless higher-order topological phase \cite{PhysRevB.103.085408}. The possibility of a disorder-induced gapless higher-order topological insulator in interacting systems remains an open question for future investigation.

\section{Conclusion}\label{section4}

We have investigated two-body higher-order topological phases in a time-periodic extended Bose-Hubbard model on a square lattice. Our study reveals a rich interplay between interactions and Floquet engineering, leading to both normal and anomalous Floquet second-order topological corner states of doublons (i.e., bound boson pairs). In the high-frequency approximation and strong-interaction limit, we derive an effective doublon Hamiltonian, demonstrating that the periodically driven interacting lattice maps onto a generalized Benalcazar-Bernevig-Hughes (BBH) model \cite{Benalcazar61}. Furthermore, we show that these interaction-induced topological phases are robust against disorder, confirming their stability. The proposed model is experimentally feasible across various platforms, including superconducting qubit arrays \cite{Roushan2016,PhysRevLett.128.213903}, ultracold atomic systems \cite{PhysRevX.7.031057,RevModPhys.91.015005}, and digital quantum computers \cite{PhysRevLett.129.140502}. Additionally, it can be simulated using electrical circuits \cite{olekhno2020topological}. Future research will explore the effects of many-body interactions beyond two interacting bosons within the Floquet framework. Another intriguing direction is the realization of interaction-driven higher-order topological Weyl semimetal phases of doublons.

\begin{acknowledgments}
	T.L. acknowledges the support from National Natural
	Science Foundation of China (Grant No.~12274142), Introduced Innovative Team Project of Guangdong Pearl River Talents Program (Grant	No.~2021ZT09Z109), the Fundamental Research Funds for the Central Universities (Grant No.~2023ZYGXZR020),   and the Startup Grant of South China University of Technology (Grant No.~20210012).
\end{acknowledgments}

\appendix
\section{Effective Hamiltonian in high-frequency approximation and strong-interaction limit} \label{AppendixA}

When the driving frequency $\omega = 2\pi/T$ significantly exceeds all other characteristic energy scales of the system,  the effective Hamiltonian $\mathcal{H}_\textrm{FM}$ can be obtained using the Floquet-Magnus expansion \cite{Bukov2015, Eckardt2015, PhysRevB.93.144307,PhysRevB.98.104303}. In the zeroth-order approximation,  the time-independent effective Hamiltonian $\mathcal{H}_\text{FM}$ is expressed as
\begin{align}\label{HFM}
	\mathcal{H}_\text{FM} = \frac{1}{T}\int_{0}^{T} \mathcal{H}(t) \,dt=(2\mathcal{H}_{0}+\mathcal{H}_{x}+\mathcal{H}_{y})/4.
\end{align}

When the driving frequency $\omega = 2\pi/T$ is large, and the onsite interaction strength satisfies $\abs{U} \gg \abs{P}, J$, the doublon bands become well-separated from the scattering bands, as illustrated in Fig.~\ref{fig3}. In this strong-interaction limit, we further derive the effective Hamiltonian $\mathcal{H}_\textrm{eff}$ of doublons by utilizing quasi-degenerate second-order perturbation theory \cite{Bir1974,CCohenTannoudji1Atom}. To go further, we rewrite the Hamiltonian $\mathcal{H}_\text{FM}$ in Eq.~(\ref{HFM}) into $\mathcal{H}_\text{FM} = \mathcal{H}_U + \mathcal{V}$, where the onsite-interaction part $\mathcal{H}_U$ reads
\begin{align}\label{HU}
	\mathcal{H}_U   =U \sum_{x, y} \sum_{\alpha \in\{1,2,3,4\}}  n_{x, y, \alpha} n_{x, y, \alpha},           
\end{align}
and the other hopping term $\mathcal{V}$  is 
\begin{widetext}
	\begin{align}\label{V}
		\mathcal{V} =&  ~~  \frac{J}{2} \sum_{x, y}\left(a^\dagger_{x, y, 1} a_{x,y,2} + a^\dagger_{x, y,2} a_{x, y,3} + a^\dagger_{x, y,3} a_{x,y,4} - a^\dagger_{x, y,4} a_{x, y, 1} + \textrm{H.c.}\right) \nonumber \\
		& + \frac{J}{4} \sum_{x,y} \left(a^\dagger_{x+1,y,1} a_{x, y,2} + a^\dagger_{x+1,y,4} a_{x, y,3}- a^\dagger_{x,y+1,1} a_{x, y,4} + a^\dagger_{x,y+1,2} a_{x, y,3}+ \textrm{H.c.} \right)       \nonumber \\ 
		& + \frac{P}{8} \sum_{x, y} \left(a^\dagger_{x+1,y,1} a^\dagger_{x+1,y,1} a_{x, y,2} a_{x,y,2}+a^\dagger_{x+1,y, 4} a^\dagger_{x+1,y,4} a_{x,y,3} a_{x, y,3} + \text{H.c.} \right)  \nonumber \\ 
		& + \frac{P}{8} \sum_{x, y} \left( a^\dagger_{x,y+1,2} a^\dagger_{x,y+1,2} a_{x, y,3} a_{x, y,3} - a^\dagger_{x,y+1,1} a^\dagger_{x,y+1,1} a_{x, y,4} a_{x,y,4} + \textrm{H.c.}\right),            
	\end{align}
\end{widetext}
where we treat the hopping part $\mathcal{V}$ in Eq.~(\ref{V}) as a perturbation in the strong-interaction regime with $\abs{U} \gg \abs{P}, J$.

Furthermore, in the strong-interaction limit, two bosons are tightly bound to occupy the same site. These states can be thus described as a set of doubly occupied states as
\begin{align}\label{doubleoccupiedstates}
	\ket{d_{x,y,\alpha}}= d^\dagger_{x,y,\alpha} \ket{0} /\sqrt{2} \equiv a^\dagger_{x,y,\alpha}  a^\dagger_{x,y,\alpha} \ket{0}/\sqrt{2}, 
\end{align}
where $\alpha \in \{1,2,3,4\}$. They are eigenstates of $\mathcal{H}_U$ with eigenenergy $E_d = 4U$. 

All the other two-boson states are given by
\begin{align}\label{otheroccupiedstates}
	\ket{s_{x^\prime x,y^\prime y,\alpha^\prime \alpha}} \equiv a^\dagger_{x^\prime,y^\prime,\alpha^\prime}  a^\dagger_{x,y,\alpha} \ket{0},
\end{align}
where $\alpha,\alpha^\prime \in \{1,2,3,4\}$, and requiring  $x^\prime\neq x$, $y^\prime\neq y$ or $\alpha^\prime\neq \alpha$. They are also eigenstates of $\mathcal{H}_U$ with eigenenergy $E_s = 2U$.

Based on the quasi-degenerate second-order perturbation theory, the non-zero matrix elements of the effective Hamiltonian $\mathcal{H}_\textrm{eff}$ are determined by 
\begin{align}\label{perturbationtheory}
	\bra{d} \mathcal{H}_\text{eff} \ket{d^\prime}   = & ~E_d \delta_{dd'} + \bra{d} \mathcal{V} \ket{d^\prime} \nonumber \\ 
	& + \frac{1}{2} \sum_{s} \bra{d} \mathcal{V} \ket{s} \bra{s} \mathcal{V} \ket{d^\prime} \nonumber \\ 
	& \times\left[\frac{1}{E_d-E_s} + \frac{1}{E_{d^\prime}-E_s}\right].         
\end{align}

The matrix elements in Eq.~(\ref{perturbationtheory}) determine the effective doublon onsite energy when $\ket{d} = \ket{d^\prime}$ and and the effective doublon coupling strength for $\ket{d} \neq \ket{d^\prime}$. Inserting Eqs.~(\ref{HU}-\ref{otheroccupiedstates}) into Eq.~(\ref{perturbationtheory}),  the matrix elements are calculated as
\begin{align}\label{eq1}
	\bra{d_{x,y,1}} \mathcal{H}_\text{eff} \ket{d_{x,y,2}} & = \bra{d_{x,y,2}} \mathcal{H}_\text{eff} \ket{d_{x,y,3}} \nonumber \\ & = \bra{d_{x,y,3}} \mathcal{H}_\text{eff} \ket{d_{x,y,4}} \nonumber \\ &  = \bra{d_{x,y,4}} \mathcal{H}_\text{eff} \ket{d_{x,y,1}} \nonumber \\ & = \frac{J^2}{4U},
\end{align}
\begin{align}\label{eq2}
	\bra{d_{x+1,y,1}} \mathcal{H}_\text{eff} \ket{d_{x,y,2}}  & = \bra{d_{x,y+1,2}} \mathcal{H}_\text{eff} \ket{d_{x,y,3}} \nonumber \\ &  = \bra{d_{x+1,y,4}} \mathcal{H}_\text{eff} \ket{d_{x,y,3}} \nonumber \\ & = \frac{J^2}{16U} + \frac{P}{4},
\end{align}
\begin{align}\label{eq3}
	\bra{d_{x,y+1,1}} \mathcal{H}_\text{eff} \ket{d_{x,y,4}} = \frac{J^2}{16U} - \frac{P}{4},
\end{align}
and
\begin{align}\label{eq4}
	\bra{d_{x,y,1}} \mathcal{H}_\text{eff} \ket{d_{x,y,1}} & = \bra{d_{x,y,2}} \mathcal{H}_\text{eff} \ket{d_{x,y,2}} \nonumber \\ &  = \bra{d_{x,y,3}} \mathcal{H}_\text{eff} \ket{d_{x,y,3}} \nonumber \\ &  = \bra{d_{x,y,4}} \mathcal{H}_\text{eff} \ket{d_{x,y,4}} \nonumber \\ &  = 4U + \frac{5J^2}{8U}.
\end{align}

For the effective doublon onsite energy, the equation (\ref{eq4}) is only valid for PBCs. In a finite lattice with OBCs, the onsite energy varies depending on the site location: it is $4U + 9J^2/(16U)$ along the finite edges, $4U + J^2/(2U)$  at the corners, and $4U + 5J^2/(8U)$  at the bulk sites. Thus, under open boundary conditions, the renormalized onsite potentials differ between bulk and boundary sites.

According to the above derivations, the effective Hamiltonian $\mathcal{H}_\text{eff}$ of doublons in the high-frequency approximation and strong-interaction limit can be expressed as
\begin{widetext}
\begin{align}\label{Heff22}
	\mathcal{H}_\text{eff}   =&  ~   \frac{J^2}{4U} \sum_{x, y}\left(d^\dagger_{x, y, 1} d_{x,y,2} + d^\dagger_{x, y,2} d_{x, y,3} + d^\dagger_{x, y,3} d_{x,y,4} + d^\dagger_{x, y,4} d_{x, y, 1} + \textrm{H.c.}\right) \nonumber \\
	& + \frac{J^2}{16U} \sum_{x,y} \left(d^\dagger_{x+1,y,1} d_{x, y,2} + d^\dagger_{x+1,y,4} d_{x, y,3}+ d^\dagger_{x,y+1,1} d_{x, y,4} + d^\dagger_{x,y+1,2} d_{x, y,3}+ \textrm{H.c.} \right)       \nonumber \\ 
	& + \frac{P}{4} \sum_{x,y} \left(d^\dagger_{x+1,y,1} d_{x, y,2} + d^\dagger_{x+1,y,4} d_{x, y,3}- d^\dagger_{x,y+1,1} d_{x, y,4} + d^\dagger_{x,y+1,2} d_{x, y,3}+ \textrm{H.c.} \right)       \nonumber \\
	& +  \sum_{x, y} U_\text{eff}(x,y) \left(n_{x, y, 1} n_{x, y, 1}  + n_{x, y, 2} n_{x, y, 2} + n_{x, y, 3} n_{x, y, 3} + n_{x, y, 4} n_{x, y, 4} \right),           
\end{align}
\end{widetext}
where $U_\text{eff}(x,y) = 4U + 5J^2/(8U)$ under the PBCs.  Under the OBCs, $U_\text{eff}(x,y)$ depends on whether $(x,y)$ is in the bulk, on the edge, or at the corner under the OBCs, and can be expressed as
\begin{align}\label{Ueff22}
	U_\text{eff}(x,y) = \begin{cases}
		4U + \frac{5J^2}{8U}, &~\text{if}~ (x,y) ~\text{in the bulk}, \\
		4U + \frac{9J^2}{16U}, &~\text{if}~ (x,y) ~\text{on the edge}, \\
		4U + \frac{J^2}{2U}, &~\text{if}~ (x,y) ~\text{at the corner}.
	\end{cases} 
\end{align}

According to Eq.~(\ref{Heff22}), in the high-frequency approximation and strong-interaction limit, the effective Hamiltonian $\mathcal{H}_\text{eff}$ describes a doublon as a quasiparticle on a square lattice. This lattice realizes a generalized Benalcazar-Bernevig-Hughes (BBH) model \cite{Benalcazar61, PhysRevB.96.245115}, featuring dimerized hoppings along both  the $x$ and $y$ directions, analogous to the one-dimensional Su-Schrieffer-Heeger (SSH) model.  Additionally, a  $\pi$-flux is introduced in alternating plaquettes due to the negative sign in the third term of the effective Hamiltonian $\mathcal{H}_\text{eff}$.    As a result, the effective Hamiltonian supports topological second-order corner states.

\begin{figure}[tb]
	\centering
	\includegraphics[width=8.6cm]{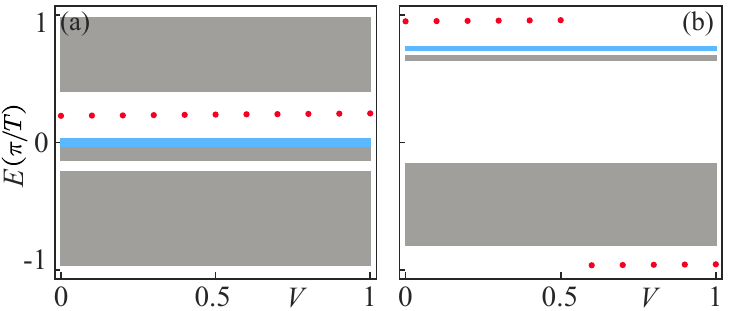}
	\caption{ (a,b) The eigenenergies of the Floquet Hamiltonian under  OBCs in both the   $x$ and $y$ direction, with a compensation potential $V$ applied at the corner and edge sites. The filled regions represent bulk (black) and edge (blue) states, while red dots indicate in-gap corner states. The parameters are (a) $J = 1$, $P = 3$, $U  = 8$, $T=1$, and (b) $J=0.5$, $P = 3$, $U  = 2.35$, $T=1$.}
	\label{FigS2}
\end{figure}

\section{Effects of Tamm-Shockley States}\label{AppendixB}	

As shown in Eqs.~(\ref{Heff22}) and (\ref{Ueff22}), under OBCs, the renormalized onsite potentials differ between boundary and bulk sites. This discrepancy in onsite potentials can give rise to two-particle Tamm-Shockley states \cite{Shockley1932,PhysRev.56.317}, which are localized at the corners and have no topological origin. In this section, we demonstrate that the in-gap states localized at the corners are not Tamm-Shockley states, but rather arise from the topological nature of the bulk bands.

To eliminate the effects of the Tamm-Shockley states, we introduce an additional compensation potential $V$ to remove the onsite potential differences between the bulk and corner sites. The compensated Hamiltonian reads $\mathcal{H}_\textrm{com}(t) = \mathcal{H}(t) + V \sum_{x_c, y_c,\alpha} n_{x_c,y_c,\alpha}$ ($\alpha \in \{1,2,3,4\}$), where $(x_c,~y_c)$  refers to    the corner and edge sites.

Figure \ref{FigS2} the eigenenergies of the Floquet Hamiltonian under  OBCs in both the   $x$ and $y$ direction, with a compensation potential $V$ applied at the corner and edge sites.  As the compensation potential $V$ increases, the band gap of the bulk bands remains open [see Fig.~\ref{FigS2}(a,b)]. Therefore, the introduction of a finite compensation potential $V$ does not induce a topological phase transition.  This potential effectively offsets the energy difference between bulk and boundary sites, which arises due to onsite boson-boson interactions in the presence of open boundaries. Furthermore, for both the normal and anomalous boundary states, the in-gap corner states do not merge into the bulk states, indicating that they are not Tamm-Shockley states but instead result from the topological nature of the bulk bands.


\begin{thebibliography}{102}%
	\makeatletter
	\providecommand \@ifxundefined [1]{%
		\@ifx{#1\undefined}
	}%
	\providecommand \@ifnum [1]{%
		\ifnum #1\expandafter \@firstoftwo
		\else \expandafter \@secondoftwo
		\fi
	}%
	\providecommand \@ifx [1]{%
		\ifx #1\expandafter \@firstoftwo
		\else \expandafter \@secondoftwo
		\fi
	}%
	\providecommand \natexlab [1]{#1}%
	\providecommand \enquote  [1]{``#1''}%
	\providecommand \bibnamefont  [1]{#1}%
	\providecommand \bibfnamefont [1]{#1}%
	\providecommand \citenamefont [1]{#1}%
	\providecommand \href@noop [0]{\@secondoftwo}%
	\providecommand \href [0]{\begingroup \@sanitize@url \@href}%
	\providecommand \@href[1]{\@@startlink{#1}\@@href}%
	\providecommand \@@href[1]{\endgroup#1\@@endlink}%
	\providecommand \@sanitize@url [0]{\catcode `\\12\catcode `\$12\catcode
		`\&12\catcode `\#12\catcode `\^12\catcode `\_12\catcode `\%12\relax}%
	\providecommand \@@startlink[1]{}%
	\providecommand \@@endlink[0]{}%
	\providecommand \url  [0]{\begingroup\@sanitize@url \@url }%
	\providecommand \@url [1]{\endgroup\@href {#1}{\urlprefix }}%
	\providecommand \urlprefix  [0]{URL }%
	\providecommand \Eprint [0]{\href }%
	\providecommand \doibase [0]{http://dx.doi.org/}%
	\providecommand \selectlanguage [0]{\@gobble}%
	\providecommand \bibinfo  [0]{\@secondoftwo}%
	\providecommand \bibfield  [0]{\@secondoftwo}%
	\providecommand \translation [1]{[#1]}%
	\providecommand \BibitemOpen [0]{}%
	\providecommand \bibitemStop [0]{}%
	\providecommand \bibitemNoStop [0]{.\EOS\space}%
	\providecommand \EOS [0]{\spacefactor3000\relax}%
	\providecommand \BibitemShut  [1]{\csname bibitem#1\endcsname}%
	\let\auto@bib@innerbib\@empty
	\bibitem [{\citenamefont {Zhang}\ \emph {et~al.}(2013)\citenamefont {Zhang},
		\citenamefont {Kane},\ and\ \citenamefont {Mele}}]{PhysRevLett.110.046404}%
	\BibitemOpen
	\bibfield  {author} {\bibinfo {author} {\bibfnamefont {F.}~\bibnamefont
			{Zhang}}, \bibinfo {author} {\bibfnamefont {C.~L.}\ \bibnamefont {Kane}}, \
		and\ \bibinfo {author} {\bibfnamefont {E.~J.}\ \bibnamefont {Mele}},\
	}\bibfield  {title} {\enquote {\bibinfo {title} {Surface state magnetization
				and chiral edge states on topological insulators},}\ }\href
	{https://link.aps.org/doi/10.1103/PhysRevLett.110.046404} {\bibfield
		{journal} {\bibinfo  {journal} {Phys. Rev. Lett.}\ }\textbf {\bibinfo
			{volume} {110}},\ \bibinfo {pages} {046404} (\bibinfo {year}
		{2013})}\BibitemShut {NoStop}%
	\bibitem [{\citenamefont {Benalcazar}\ \emph
		{et~al.}(2017{\natexlab{a}})\citenamefont {Benalcazar}, \citenamefont
		{Bernevig},\ and\ \citenamefont {Hughes}}]{Benalcazar61}%
	\BibitemOpen
	\bibfield  {author} {\bibinfo {author} {\bibfnamefont {W.~A.}\ \bibnamefont
			{Benalcazar}}, \bibinfo {author} {\bibfnamefont {B.~A.}\ \bibnamefont
			{Bernevig}}, \ and\ \bibinfo {author} {\bibfnamefont {T.~L.}\ \bibnamefont
			{Hughes}},\ }\bibfield  {title} {\enquote {\bibinfo {title} {Quantized
				electric multipole insulators},}\ }\href
	{https://www.science.org/doi/10.1126/science.aah6442} {\bibfield  {journal}
		{\bibinfo  {journal} {Science}\ }\textbf {\bibinfo {volume} {357}},\ \bibinfo
		{pages} {61} (\bibinfo {year} {2017}{\natexlab{a}})}\BibitemShut {NoStop}%
	\bibitem [{\citenamefont {Benalcazar}\ \emph
		{et~al.}(2017{\natexlab{b}})\citenamefont {Benalcazar}, \citenamefont
		{Bernevig},\ and\ \citenamefont {Hughes}}]{PhysRevB.96.245115}%
	\BibitemOpen
	\bibfield  {author} {\bibinfo {author} {\bibfnamefont {W.~A.}\ \bibnamefont
			{Benalcazar}}, \bibinfo {author} {\bibfnamefont {B.~A.}\ \bibnamefont
			{Bernevig}}, \ and\ \bibinfo {author} {\bibfnamefont {T.~L.}\ \bibnamefont
			{Hughes}},\ }\bibfield  {title} {\enquote {\bibinfo {title} {Electric
				multipole moments, topological multipole moment pumping, and chiral hinge
				states in crystalline insulators},}\ }\href
	{https://link.aps.org/doi/10.1103/PhysRevB.96.245115} {\bibfield  {journal}
		{\bibinfo  {journal} {Phys. Rev. B}\ }\textbf {\bibinfo {volume} {96}},\
		\bibinfo {pages} {245115} (\bibinfo {year} {2017}{\natexlab{b}})}\BibitemShut
	{NoStop}%
	\bibitem [{\citenamefont {Langbehn}\ \emph {et~al.}(2017)\citenamefont
		{Langbehn}, \citenamefont {Peng}, \citenamefont {Trifunovic}, \citenamefont
		{von Oppen},\ and\ \citenamefont {Brouwer}}]{PhysRevLett.119.246401}%
	\BibitemOpen
	\bibfield  {author} {\bibinfo {author} {\bibfnamefont {J.}~\bibnamefont
			{Langbehn}}, \bibinfo {author} {\bibfnamefont {Y.}~\bibnamefont {Peng}},
		\bibinfo {author} {\bibfnamefont {L.}~\bibnamefont {Trifunovic}}, \bibinfo
		{author} {\bibfnamefont {F.}~\bibnamefont {von Oppen}}, \ and\ \bibinfo
		{author} {\bibfnamefont {P.~W.}\ \bibnamefont {Brouwer}},\ }\bibfield
	{title} {\enquote {\bibinfo {title} {Reflection-symmetric second-order
				topological insulators and superconductors},}\ }\href
	{https://link.aps.org/doi/10.1103/PhysRevLett.119.246401} {\bibfield
		{journal} {\bibinfo  {journal} {Phys. Rev. Lett.}\ }\textbf {\bibinfo
			{volume} {119}},\ \bibinfo {pages} {246401} (\bibinfo {year}
		{2017})}\BibitemShut {NoStop}%
	\bibitem [{\citenamefont {Song}\ \emph {et~al.}(2017)\citenamefont {Song},
		\citenamefont {Fang},\ and\ \citenamefont {Fang}}]{PhysRevLett.119.246402}%
	\BibitemOpen
	\bibfield  {author} {\bibinfo {author} {\bibfnamefont {Z.}~\bibnamefont
			{Song}}, \bibinfo {author} {\bibfnamefont {Z.}~\bibnamefont {Fang}}, \ and\
		\bibinfo {author} {\bibfnamefont {C.}~\bibnamefont {Fang}},\ }\bibfield
	{title} {\enquote {\bibinfo {title} {$(d\ensuremath{-}2)$-dimensional edge
				states of rotation symmetry protected topological states},}\ }\href
	{https://link.aps.org/doi/10.1103/PhysRevLett.119.246401} {\bibfield
		{journal} {\bibinfo  {journal} {Phys. Rev. Lett.}\ }\textbf {\bibinfo
			{volume} {119}},\ \bibinfo {pages} {246402} (\bibinfo {year}
		{2017})}\BibitemShut {NoStop}%
	\bibitem [{\citenamefont {Xie}\ \emph {et~al.}(2021)\citenamefont {Xie},
		\citenamefont {Wang}, \citenamefont {Zhang}, \citenamefont {Zhan},
		\citenamefont {Jiang}, \citenamefont {Lu},\ and\ \citenamefont
		{Chen}}]{Xie2021}%
	\BibitemOpen
	\bibfield  {author} {\bibinfo {author} {\bibfnamefont {B.}~\bibnamefont
			{Xie}}, \bibinfo {author} {\bibfnamefont {H.-X.}\ \bibnamefont {Wang}},
		\bibinfo {author} {\bibfnamefont {X.}~\bibnamefont {Zhang}}, \bibinfo
		{author} {\bibfnamefont {P.}~\bibnamefont {Zhan}}, \bibinfo {author}
		{\bibfnamefont {J.-H.}\ \bibnamefont {Jiang}}, \bibinfo {author}
		{\bibfnamefont {M.}~\bibnamefont {Lu}}, \ and\ \bibinfo {author}
		{\bibfnamefont {Y.}~\bibnamefont {Chen}},\ }\bibfield  {title} {\enquote
		{\bibinfo {title} {Higher-order band topology},}\ }\href {\doibase
		10.1038/s42254-021-00323-4} {\bibfield  {journal} {\bibinfo  {journal} {Nat.
				Rev. Phys.}\ }\textbf {\bibinfo {volume} {3}},\ \bibinfo {pages} {520}
		(\bibinfo {year} {2021})}\BibitemShut {NoStop}%
	\bibitem [{\citenamefont {Peterson}\ \emph {et~al.}(2018)\citenamefont
		{Peterson}, \citenamefont {Benalcazar}, \citenamefont {Hughes},\ and\
		\citenamefont {Bahl}}]{Peterson2018}%
	\BibitemOpen
	\bibfield  {author} {\bibinfo {author} {\bibfnamefont {C.~W.}\ \bibnamefont
			{Peterson}}, \bibinfo {author} {\bibfnamefont {W.~A.}\ \bibnamefont
			{Benalcazar}}, \bibinfo {author} {\bibfnamefont {T.~L.}\ \bibnamefont
			{Hughes}}, \ and\ \bibinfo {author} {\bibfnamefont {G.}~\bibnamefont
			{Bahl}},\ }\bibfield  {title} {\enquote {\bibinfo {title} {A quantized
				microwave quadrupole insulator with topologically protected corner states},}\
	}\href {http://dx.doi.org/10.1038/nature25777 http://10.0.4.14/nature25777}
	{\bibfield  {journal} {\bibinfo  {journal} {Nature}\ }\textbf {\bibinfo
			{volume} {555}},\ \bibinfo {pages} {346} (\bibinfo {year}
		{2018})}\BibitemShut {NoStop}%
	\bibitem [{\citenamefont {Serra-Garcia}\ \emph {et~al.}(2018)\citenamefont
		{Serra-Garcia}, \citenamefont {Peri}, \citenamefont {S{\"{u}}sstrunk},
		\citenamefont {Bilal}, \citenamefont {Larsen}, \citenamefont {Villanueva},\
		and\ \citenamefont {Huber}}]{Serra-Garcia2018}%
	\BibitemOpen
	\bibfield  {author} {\bibinfo {author} {\bibfnamefont {M.}~\bibnamefont
			{Serra-Garcia}}, \bibinfo {author} {\bibfnamefont {V.}~\bibnamefont {Peri}},
		\bibinfo {author} {\bibfnamefont {R.}~\bibnamefont {S{\"{u}}sstrunk}},
		\bibinfo {author} {\bibfnamefont {O.~R.}\ \bibnamefont {Bilal}}, \bibinfo
		{author} {\bibfnamefont {T.}~\bibnamefont {Larsen}}, \bibinfo {author}
		{\bibfnamefont {L.~G.}\ \bibnamefont {Villanueva}}, \ and\ \bibinfo {author}
		{\bibfnamefont {S.~D.}\ \bibnamefont {Huber}},\ }\bibfield  {title} {\enquote
		{\bibinfo {title} {Observation of a phononic quadrupole topological
				insulator},}\ }\href {http://dx.doi.org/10.1038/nature25156} {\bibfield
		{journal} {\bibinfo  {journal} {Nature}\ }\textbf {\bibinfo {volume} {555}},\
		\bibinfo {pages} {342} (\bibinfo {year} {2018})}\BibitemShut {NoStop}%
	\bibitem [{\citenamefont {Geier}\ \emph {et~al.}(2018)\citenamefont {Geier},
		\citenamefont {Trifunovic}, \citenamefont {Hoskam},\ and\ \citenamefont
		{Brouwer}}]{arXiv:1801.10053}%
	\BibitemOpen
	\bibfield  {author} {\bibinfo {author} {\bibfnamefont {M.}~\bibnamefont
			{Geier}}, \bibinfo {author} {\bibfnamefont {L.}~\bibnamefont {Trifunovic}},
		\bibinfo {author} {\bibfnamefont {M.}~\bibnamefont {Hoskam}}, \ and\ \bibinfo
		{author} {\bibfnamefont {P.~W.}\ \bibnamefont {Brouwer}},\ }\bibfield
	{title} {\enquote {\bibinfo {title} {Second-order topological insulators and
				superconductors with an order-two crystalline symmetry},}\ }\href
	{https://link.aps.org/doi/10.1103/PhysRevB.97.205135} {\bibfield  {journal}
		{\bibinfo  {journal} {Phys. Rev. B}\ }\textbf {\bibinfo {volume} {97}},\
		\bibinfo {pages} {205135} (\bibinfo {year} {2018})}\BibitemShut {NoStop}%
	\bibitem [{\citenamefont {Ezawa}(2018)}]{PhysRevLett.120.026801}%
	\BibitemOpen
	\bibfield  {author} {\bibinfo {author} {\bibfnamefont {M.}~\bibnamefont
			{Ezawa}},\ }\bibfield  {title} {\enquote {\bibinfo {title} {Higher-order
				topological insulators and semimetals on the breathing kagome and pyrochlore
				lattices},}\ }\href {https://link.aps.org/doi/10.1103/PhysRevLett.120.026801}
	{\bibfield  {journal} {\bibinfo  {journal} {Phys. Rev. Lett.}\ }\textbf
		{\bibinfo {volume} {120}},\ \bibinfo {pages} {026801} (\bibinfo {year}
		{2018})}\BibitemShut {NoStop}%
	\bibitem [{\citenamefont {Schindler}\ \emph
		{et~al.}(2018{\natexlab{a}})\citenamefont {Schindler}, \citenamefont {Cook},
		\citenamefont {Vergniory}, \citenamefont {Wang}, \citenamefont {Parkin},
		\citenamefont {Bernevig},\ and\ \citenamefont {Neupert}}]{TitusSciAdv2018}%
	\BibitemOpen
	\bibfield  {author} {\bibinfo {author} {\bibfnamefont {F.}~\bibnamefont
			{Schindler}}, \bibinfo {author} {\bibfnamefont {A.~M.}\ \bibnamefont {Cook}},
		\bibinfo {author} {\bibfnamefont {M.~G.}\ \bibnamefont {Vergniory}}, \bibinfo
		{author} {\bibfnamefont {Z.}~\bibnamefont {Wang}}, \bibinfo {author}
		{\bibfnamefont {S.~S.~P.}\ \bibnamefont {Parkin}}, \bibinfo {author}
		{\bibfnamefont {B.~A.}\ \bibnamefont {Bernevig}}, \ and\ \bibinfo {author}
		{\bibfnamefont {T.}~\bibnamefont {Neupert}},\ }\bibfield  {title} {\enquote
		{\bibinfo {title} {Higher-order topological insulators},}\ }\href
	{http://advances.sciencemag.org/content/4/6/eaat0346} {\bibfield  {journal}
		{\bibinfo  {journal} {Sci. Adv.}\ }\textbf {\bibinfo {volume} {4}},\ \bibinfo
		{pages} {eaat0346} (\bibinfo {year} {2018}{\natexlab{a}})}\BibitemShut
	{NoStop}%
	\bibitem [{\citenamefont {Ni}\ \emph {et~al.}(2018)\citenamefont {Ni},
		\citenamefont {Weiner}, \citenamefont {Al{\`{u}}},\ and\ \citenamefont
		{Khanikaev}}]{Ni2018}%
	\BibitemOpen
	\bibfield  {author} {\bibinfo {author} {\bibfnamefont {X.}~\bibnamefont
			{Ni}}, \bibinfo {author} {\bibfnamefont {M.}~\bibnamefont {Weiner}}, \bibinfo
		{author} {\bibfnamefont {A.}~\bibnamefont {Al{\`{u}}}}, \ and\ \bibinfo
		{author} {\bibfnamefont {A.~B.}\ \bibnamefont {Khanikaev}},\ }\bibfield
	{title} {\enquote {\bibinfo {title} {Observation of higher-order topological
				acoustic states protected by generalized chiral symmetry},}\ }\href {\doibase
		10.1038/s41563-018-0252-9} {\bibfield  {journal} {\bibinfo  {journal} {Nat.
				Mater.}\ }\textbf {\bibinfo {volume} {18}},\ \bibinfo {pages} {113} (\bibinfo
		{year} {2018})}\BibitemShut {NoStop}%
	\bibitem [{\citenamefont {Xue}\ \emph {et~al.}(2018)\citenamefont {Xue},
		\citenamefont {Yang}, \citenamefont {Gao}, \citenamefont {Chong},\ and\
		\citenamefont {Zhang}}]{Xue2018}%
	\BibitemOpen
	\bibfield  {author} {\bibinfo {author} {\bibfnamefont {H.}~\bibnamefont
			{Xue}}, \bibinfo {author} {\bibfnamefont {Y.}~\bibnamefont {Yang}}, \bibinfo
		{author} {\bibfnamefont {F.}~\bibnamefont {Gao}}, \bibinfo {author}
		{\bibfnamefont {Y.}~\bibnamefont {Chong}}, \ and\ \bibinfo {author}
		{\bibfnamefont {B.}~\bibnamefont {Zhang}},\ }\bibfield  {title} {\enquote
		{\bibinfo {title} {Acoustic higher-order topological insulator on a kagome
				lattice},}\ }\href {\doibase 10.1038/s41563-018-0251-x} {\bibfield  {journal}
		{\bibinfo  {journal} {Nat. Mater.}\ }\textbf {\bibinfo {volume} {18}},\
		\bibinfo {pages} {108} (\bibinfo {year} {2018})}\BibitemShut {NoStop}%
	\bibitem [{\citenamefont {Khalaf}(2018)}]{arXiv:1801.10050}%
	\BibitemOpen
	\bibfield  {author} {\bibinfo {author} {\bibfnamefont {E.}~\bibnamefont
			{Khalaf}},\ }\bibfield  {title} {\enquote {\bibinfo {title} {Higher-order
				topological insulators and superconductors protected by inversion
				symmetry},}\ }\href {https://link.aps.org/doi/10.1103/PhysRevB.97.205136}
	{\bibfield  {journal} {\bibinfo  {journal} {Phys. Rev. B}\ }\textbf {\bibinfo
			{volume} {97}},\ \bibinfo {pages} {205136} (\bibinfo {year}
		{2018})}\BibitemShut {NoStop}%
	\bibitem [{\citenamefont {Schindler}\ \emph
		{et~al.}(2018{\natexlab{b}})\citenamefont {Schindler}, \citenamefont {Wang},
		\citenamefont {Vergniory}, \citenamefont {Cook}, \citenamefont {Murani},
		\citenamefont {Sengupta}, \citenamefont {Kasumov}, \citenamefont {Deblock},
		\citenamefont {Drozdov}, \citenamefont {Bouchiat}, \citenamefont {Guéron},
		\citenamefont {Yazdani}, \citenamefont {Bernevig},\ and\ \citenamefont
		{Neupert}}]{arXiv:1802.02585}%
	\BibitemOpen
	\bibfield  {author} {\bibinfo {author} {\bibfnamefont {F.}~\bibnamefont
			{Schindler}}, \bibinfo {author} {\bibfnamefont {Z.}~\bibnamefont {Wang}},
		\bibinfo {author} {\bibfnamefont {M.~G.}\ \bibnamefont {Vergniory}}, \bibinfo
		{author} {\bibfnamefont {A.~M.}\ \bibnamefont {Cook}}, \bibinfo {author}
		{\bibfnamefont {A.}~\bibnamefont {Murani}}, \bibinfo {author} {\bibfnamefont
			{S.}~\bibnamefont {Sengupta}}, \bibinfo {author} {\bibfnamefont {A.~Y.}\
			\bibnamefont {Kasumov}}, \bibinfo {author} {\bibfnamefont {R.}~\bibnamefont
			{Deblock}}, \bibinfo {author} {\bibfnamefont {S.~J.~I.}\ \bibnamefont
			{Drozdov}}, \bibinfo {author} {\bibfnamefont {H.}~\bibnamefont {Bouchiat}},
		\bibinfo {author} {\bibfnamefont {S.}~\bibnamefont {Guéron}}, \bibinfo
		{author} {\bibfnamefont {A.}~\bibnamefont {Yazdani}}, \bibinfo {author}
		{\bibfnamefont {B.~A.}\ \bibnamefont {Bernevig}}, \ and\ \bibinfo {author}
		{\bibfnamefont {T.}~\bibnamefont {Neupert}},\ }\bibfield  {title} {\enquote
		{\bibinfo {title} {Higher-order topology in \uppercase{B}ismuth},}\ }\href
	{https://www.nature.com/articles/s41567-018-0224-7} {\bibfield  {journal}
		{\bibinfo  {journal} {Nat. Phys.}\ }\textbf {\bibinfo {volume} {14}},\
		\bibinfo {pages} {918} (\bibinfo {year} {2018}{\natexlab{b}})}\BibitemShut
	{NoStop}%
	\bibitem [{\citenamefont {Park}\ \emph {et~al.}(2019)\citenamefont {Park},
		\citenamefont {Kim}, \citenamefont {Cho},\ and\ \citenamefont
		{Lee}}]{PhysRevLett.123.216803}%
	\BibitemOpen
	\bibfield  {author} {\bibinfo {author} {\bibfnamefont {M.~J.}\ \bibnamefont
			{Park}}, \bibinfo {author} {\bibfnamefont {Y.}~\bibnamefont {Kim}}, \bibinfo
		{author} {\bibfnamefont {G.~Y.}\ \bibnamefont {Cho}}, \ and\ \bibinfo
		{author} {\bibfnamefont {S.}~\bibnamefont {Lee}},\ }\bibfield  {title}
	{\enquote {\bibinfo {title} {Higher-order topological insulator in twisted
				bilayer graphene},}\ }\href {\doibase 10.1103/PhysRevLett.123.216803}
	{\bibfield  {journal} {\bibinfo  {journal} {Phys. Rev. Lett.}\ }\textbf
		{\bibinfo {volume} {123}},\ \bibinfo {pages} {216803} (\bibinfo {year}
		{2019})}\BibitemShut {NoStop}%
	\bibitem [{\citenamefont {Mittal}\ \emph {et~al.}(2019)\citenamefont {Mittal},
		\citenamefont {Orre}, \citenamefont {Zhu}, \citenamefont {Gorlach},
		\citenamefont {Poddubny},\ and\ \citenamefont {Hafezi}}]{Mittal2019}%
	\BibitemOpen
	\bibfield  {author} {\bibinfo {author} {\bibfnamefont {S.}~\bibnamefont
			{Mittal}}, \bibinfo {author} {\bibfnamefont {V.~Vikram}\ \bibnamefont
			{Orre}}, \bibinfo {author} {\bibfnamefont {G.}~\bibnamefont {Zhu}}, \bibinfo
		{author} {\bibfnamefont {M.~A.}\ \bibnamefont {Gorlach}}, \bibinfo {author}
		{\bibfnamefont {A.}~\bibnamefont {Poddubny}}, \ and\ \bibinfo {author}
		{\bibfnamefont {M.}~\bibnamefont {Hafezi}},\ }\bibfield  {title} {\enquote
		{\bibinfo {title} {Photonic quadrupole topological phases},}\ }\href
	{\doibase 10.1038/s41566-019-0452-0} {\bibfield  {journal} {\bibinfo
			{journal} {Nat. Photon.}\ }\textbf {\bibinfo {volume} {13}},\ \bibinfo
		{pages} {692} (\bibinfo {year} {2019})}\BibitemShut {NoStop}%
	\bibitem [{\citenamefont {Hassan}\ \emph {et~al.}(2019)\citenamefont {Hassan},
		\citenamefont {Kunst}, \citenamefont {Moritz}, \citenamefont {Andler},
		\citenamefont {Bergholtz},\ and\ \citenamefont {Bourennane}}]{ElHassan2019}%
	\BibitemOpen
	\bibfield  {author} {\bibinfo {author} {\bibfnamefont {A.~El}\ \bibnamefont
			{Hassan}}, \bibinfo {author} {\bibfnamefont {F.~K.}\ \bibnamefont {Kunst}},
		\bibinfo {author} {\bibfnamefont {A.}~\bibnamefont {Moritz}}, \bibinfo
		{author} {\bibfnamefont {G.}~\bibnamefont {Andler}}, \bibinfo {author}
		{\bibfnamefont {E.~J.}\ \bibnamefont {Bergholtz}}, \ and\ \bibinfo {author}
		{\bibfnamefont {M.}~\bibnamefont {Bourennane}},\ }\bibfield  {title}
	{\enquote {\bibinfo {title} {Corner states of light in photonic
				waveguides},}\ }\href {\doibase 10.1038/s41566-019-0519-y} {\bibfield
		{journal} {\bibinfo  {journal} {Nat. Photon.}\ }\textbf {\bibinfo {volume}
			{13}},\ \bibinfo {pages} {697} (\bibinfo {year} {2019})}\BibitemShut
	{NoStop}%
	\bibitem [{\citenamefont {Zhang}\ \emph
		{et~al.}(2019{\natexlab{a}})\citenamefont {Zhang}, \citenamefont {Wang},
		\citenamefont {Lin}, \citenamefont {Tian}, \citenamefont {Xie}, \citenamefont
		{Lu}, \citenamefont {Chen},\ and\ \citenamefont {Jiang}}]{Zhang2019}%
	\BibitemOpen
	\bibfield  {author} {\bibinfo {author} {\bibfnamefont {X.}~\bibnamefont
			{Zhang}}, \bibinfo {author} {\bibfnamefont {H.~X.}\ \bibnamefont {Wang}},
		\bibinfo {author} {\bibfnamefont {Z.~K.}\ \bibnamefont {Lin}}, \bibinfo
		{author} {\bibfnamefont {Y.}~\bibnamefont {Tian}}, \bibinfo {author}
		{\bibfnamefont {B.}~\bibnamefont {Xie}}, \bibinfo {author} {\bibfnamefont
			{M.~H.}\ \bibnamefont {Lu}}, \bibinfo {author} {\bibfnamefont {Y.~F.}\
			\bibnamefont {Chen}}, \ and\ \bibinfo {author} {\bibfnamefont {J.~H.}\
			\bibnamefont {Jiang}},\ }\bibfield  {title} {\enquote {\bibinfo {title}
			{Second-order topology and multidimensional topological transitions in sonic
				crystals},}\ }\href {\doibase 10.1038/s41567-019-0472-1} {\bibfield
		{journal} {\bibinfo  {journal} {Nat. Phys.}\ }\textbf {\bibinfo {volume}
			{15}},\ \bibinfo {pages} {582} (\bibinfo {year}
		{2019}{\natexlab{a}})}\BibitemShut {NoStop}%
	\bibitem [{\citenamefont {Yang}\ \emph {et~al.}(2020)\citenamefont {Yang},
		\citenamefont {Li}, \citenamefont {Duan},\ and\ \citenamefont
		{Xu}}]{PhysRevResearch.2.033029}%
	\BibitemOpen
	\bibfield  {author} {\bibinfo {author} {\bibfnamefont {Y.~B.}\ \bibnamefont
			{Yang}}, \bibinfo {author} {\bibfnamefont {K.}~\bibnamefont {Li}}, \bibinfo
		{author} {\bibfnamefont {L.-M.}\ \bibnamefont {Duan}}, \ and\ \bibinfo
		{author} {\bibfnamefont {Y.}~\bibnamefont {Xu}},\ }\bibfield  {title}
	{\enquote {\bibinfo {title} {Type-{II} quadrupole topological insulators},}\
	}\href {\doibase 10.1103/PhysRevResearch.2.033029} {\bibfield  {journal}
		{\bibinfo  {journal} {Phys. Rev. Research}\ }\textbf {\bibinfo {volume}
			{2}},\ \bibinfo {pages} {033029} (\bibinfo {year} {2020})}\BibitemShut
	{NoStop}%
	\bibitem [{\citenamefont {Zeng}\ \emph {et~al.}(2020)\citenamefont {Zeng},
		\citenamefont {Yang},\ and\ \citenamefont {Xu}}]{PhysRevB.101.241104}%
	\BibitemOpen
	\bibfield  {author} {\bibinfo {author} {\bibfnamefont {Q.~B.}\ \bibnamefont
			{Zeng}}, \bibinfo {author} {\bibfnamefont {Y.~B.}\ \bibnamefont {Yang}}, \
		and\ \bibinfo {author} {\bibfnamefont {Y.}~\bibnamefont {Xu}},\ }\bibfield
	{title} {\enquote {\bibinfo {title} {Higher-order topological insulators and
				semimetals in generalized {A}ubry-{A}ndr\'e-{H}arper models},}\ }\href
	{\doibase 10.1103/PhysRevB.101.241104} {\bibfield  {journal} {\bibinfo
			{journal} {Phys. Rev. B}\ }\textbf {\bibinfo {volume} {101}},\ \bibinfo
		{pages} {241104} (\bibinfo {year} {2020})}\BibitemShut {NoStop}%
	\bibitem [{\citenamefont {Chen}\ \emph {et~al.}(2020)\citenamefont {Chen},
		\citenamefont {Chen}, \citenamefont {Gao}, \citenamefont {Zhou},\ and\
		\citenamefont {Xu}}]{PhysRevLett.124.036803}%
	\BibitemOpen
	\bibfield  {author} {\bibinfo {author} {\bibfnamefont {R.}~\bibnamefont
			{Chen}}, \bibinfo {author} {\bibfnamefont {C.~Z.}\ \bibnamefont {Chen}},
		\bibinfo {author} {\bibfnamefont {J.~H.}\ \bibnamefont {Gao}}, \bibinfo
		{author} {\bibfnamefont {B.}~\bibnamefont {Zhou}}, \ and\ \bibinfo {author}
		{\bibfnamefont {D.~H.}\ \bibnamefont {Xu}},\ }\bibfield  {title} {\enquote
		{\bibinfo {title} {Higher-order topological insulators in quasicrystals},}\
	}\href {\doibase 10.1103/PhysRevLett.124.036803} {\bibfield  {journal}
		{\bibinfo  {journal} {Phys. Rev. Lett.}\ }\textbf {\bibinfo {volume} {124}},\
		\bibinfo {pages} {036803} (\bibinfo {year} {2020})}\BibitemShut {NoStop}%
	\bibitem [{\citenamefont {Banerjee}\ \emph {et~al.}(2020)\citenamefont
		{Banerjee}, \citenamefont {Mandal},\ and\ \citenamefont
		{Liew}}]{PhysRevLett.124.063901}%
	\BibitemOpen
	\bibfield  {author} {\bibinfo {author} {\bibfnamefont {R.}~\bibnamefont
			{Banerjee}}, \bibinfo {author} {\bibfnamefont {S.}~\bibnamefont {Mandal}}, \
		and\ \bibinfo {author} {\bibfnamefont {T.~C.~H.}\ \bibnamefont {Liew}},\
	}\bibfield  {title} {\enquote {\bibinfo {title} {Coupling between
				exciton-polariton corner modes through edge states},}\ }\href {\doibase
		10.1103/PhysRevLett.124.063901} {\bibfield  {journal} {\bibinfo  {journal}
			{Phys. Rev. Lett.}\ }\textbf {\bibinfo {volume} {124}},\ \bibinfo {pages}
		{063901} (\bibinfo {year} {2020})}\BibitemShut {NoStop}%
	\bibitem [{\citenamefont {Zhu}(2018)}]{PRBXYZhu2018}%
	\BibitemOpen
	\bibfield  {author} {\bibinfo {author} {\bibfnamefont {X.}~\bibnamefont
			{Zhu}},\ }\bibfield  {title} {\enquote {\bibinfo {title} {Tunable
				\uppercase{M}ajorana corner states in a two-dimensional second-order
				topological superconductor induced by magnetic fields},}\ }\href
	{https://link.aps.org/doi/10.1103/PhysRevB.97.205134} {\bibfield  {journal}
		{\bibinfo  {journal} {Phys. Rev. B}\ }\textbf {\bibinfo {volume} {97}},\
		\bibinfo {pages} {205134} (\bibinfo {year} {2018})}\BibitemShut {NoStop}%
	\bibitem [{\citenamefont {Yan}\ \emph {et~al.}(2018)\citenamefont {Yan},
		\citenamefont {Song},\ and\ \citenamefont {Wang}}]{arXiv:1803.08545}%
	\BibitemOpen
	\bibfield  {author} {\bibinfo {author} {\bibfnamefont {Z.}~\bibnamefont
			{Yan}}, \bibinfo {author} {\bibfnamefont {F.}~\bibnamefont {Song}}, \ and\
		\bibinfo {author} {\bibfnamefont {Z.}~\bibnamefont {Wang}},\ }\bibfield
	{title} {\enquote {\bibinfo {title} {Majorana corner modes in a
				high-temperature platform},}\ }\href
	{https://link.aps.org/doi/10.1103/PhysRevLett.121.096803} {\bibfield
		{journal} {\bibinfo  {journal} {Phys. Rev. Lett.}\ }\textbf {\bibinfo
			{volume} {121}},\ \bibinfo {pages} {096803} (\bibinfo {year}
		{2018})}\BibitemShut {NoStop}%
	\bibitem [{\citenamefont {Liu}\ \emph {et~al.}(2018)\citenamefont {Liu},
		\citenamefont {He},\ and\ \citenamefont {Nori}}]{arXiv:1806.07002}%
	\BibitemOpen
	\bibfield  {author} {\bibinfo {author} {\bibfnamefont {T.}~\bibnamefont
			{Liu}}, \bibinfo {author} {\bibfnamefont {J.~J.}\ \bibnamefont {He}}, \ and\
		\bibinfo {author} {\bibfnamefont {F.}~\bibnamefont {Nori}},\ }\bibfield
	{title} {\enquote {\bibinfo {title} {Majorana corner states in a
				two-dimensional magnetic topological insulator on a high-temperature
				superconductor},}\ }\href
	{https://link.aps.org/doi/10.1103/PhysRevB.98.245413} {\bibfield  {journal}
		{\bibinfo  {journal} {Phys. Rev. B}\ }\textbf {\bibinfo {volume} {98}},\
		\bibinfo {pages} {245413} (\bibinfo {year} {2018})}\BibitemShut {NoStop}%
	\bibitem [{\citenamefont {Hsu}\ \emph {et~al.}(2018)\citenamefont {Hsu},
		\citenamefont {Stano}, \citenamefont {Klinovaja},\ and\ \citenamefont
		{Loss}}]{PhysRevLett.121.196801}%
	\BibitemOpen
	\bibfield  {author} {\bibinfo {author} {\bibfnamefont {C.~H.}\ \bibnamefont
			{Hsu}}, \bibinfo {author} {\bibfnamefont {P.}~\bibnamefont {Stano}}, \bibinfo
		{author} {\bibfnamefont {J.}~\bibnamefont {Klinovaja}}, \ and\ \bibinfo
		{author} {\bibfnamefont {D.}~\bibnamefont {Loss}},\ }\bibfield  {title}
	{\enquote {\bibinfo {title} {Majorana {K}ramers pairs in higher-order
				topological insulators},}\ }\href {\doibase 10.1103/PhysRevLett.121.196801}
	{\bibfield  {journal} {\bibinfo  {journal} {Phys. Rev. Lett.}\ }\textbf
		{\bibinfo {volume} {121}},\ \bibinfo {pages} {196801} (\bibinfo {year}
		{2018})}\BibitemShut {NoStop}%
	\bibitem [{\citenamefont {Yan}(2019)}]{PhysRevLett.123.177001}%
	\BibitemOpen
	\bibfield  {author} {\bibinfo {author} {\bibfnamefont {Z.}~\bibnamefont
			{Yan}},\ }\bibfield  {title} {\enquote {\bibinfo {title} {Higher-order
				topological odd-parity superconductors},}\ }\href {\doibase
		10.1103/PhysRevLett.123.177001} {\bibfield  {journal} {\bibinfo  {journal}
			{Phys. Rev. Lett.}\ }\textbf {\bibinfo {volume} {123}},\ \bibinfo {pages}
		{177001} (\bibinfo {year} {2019})}\BibitemShut {NoStop}%
	\bibitem [{\citenamefont {Zhu}(2019)}]{PhysRevLett.122.236401}%
	\BibitemOpen
	\bibfield  {author} {\bibinfo {author} {\bibfnamefont {X.}~\bibnamefont
			{Zhu}},\ }\bibfield  {title} {\enquote {\bibinfo {title} {Second-order
				topological superconductors with mixed pairing},}\ }\href {\doibase
		10.1103/PhysRevLett.122.236401} {\bibfield  {journal} {\bibinfo  {journal}
			{Phys. Rev. Lett.}\ }\textbf {\bibinfo {volume} {122}},\ \bibinfo {pages}
		{236401} (\bibinfo {year} {2019})}\BibitemShut {NoStop}%
	\bibitem [{\citenamefont {Wu}\ \emph {et~al.}(2019)\citenamefont {Wu},
		\citenamefont {Yan},\ and\ \citenamefont {Huang}}]{PhysRevB.99.020508}%
	\BibitemOpen
	\bibfield  {author} {\bibinfo {author} {\bibfnamefont {Z.}~\bibnamefont
			{Wu}}, \bibinfo {author} {\bibfnamefont {Z.}~\bibnamefont {Yan}}, \ and\
		\bibinfo {author} {\bibfnamefont {W.}~\bibnamefont {Huang}},\ }\bibfield
	{title} {\enquote {\bibinfo {title} {Higher-order topological
				superconductivity: Possible realization in {F}ermi gases and
				${\mathrm{\uppercase{s}r}}_{2}{\mathrm{\uppercase{r}u\uppercase{o}}}_{4}$},}\
	}\href {\doibase 10.1103/PhysRevB.99.020508} {\bibfield  {journal} {\bibinfo
			{journal} {Phys. Rev. B}\ }\textbf {\bibinfo {volume} {99}},\ \bibinfo
		{pages} {020508} (\bibinfo {year} {2019})}\BibitemShut {NoStop}%
	\bibitem [{\citenamefont {Liu}\ \emph {et~al.}(2019)\citenamefont {Liu},
		\citenamefont {Zhang}, \citenamefont {Ai}, \citenamefont {Gong},
		\citenamefont {Kawabata}, \citenamefont {Ueda},\ and\ \citenamefont
		{Nori}}]{PhysRevLett.122.076801}%
	\BibitemOpen
	\bibfield  {author} {\bibinfo {author} {\bibfnamefont {T.}~\bibnamefont
			{Liu}}, \bibinfo {author} {\bibfnamefont {Y.~R.}\ \bibnamefont {Zhang}},
		\bibinfo {author} {\bibfnamefont {Q.}~\bibnamefont {Ai}}, \bibinfo {author}
		{\bibfnamefont {Z.~P.}\ \bibnamefont {Gong}}, \bibinfo {author}
		{\bibfnamefont {K.}~\bibnamefont {Kawabata}}, \bibinfo {author}
		{\bibfnamefont {M.}~\bibnamefont {Ueda}}, \ and\ \bibinfo {author}
		{\bibfnamefont {F.}~\bibnamefont {Nori}},\ }\bibfield  {title} {\enquote
		{\bibinfo {title} {Second-order topological phases in non-{H}ermitian
				systems},}\ }\href {\doibase 10.1103/PhysRevLett.122.076801} {\bibfield
		{journal} {\bibinfo  {journal} {Phys. Rev. Lett.}\ }\textbf {\bibinfo
			{volume} {122}},\ \bibinfo {pages} {076801} (\bibinfo {year}
		{2019})}\BibitemShut {NoStop}%
	\bibitem [{\citenamefont {Zhang}\ \emph
		{et~al.}(2019{\natexlab{b}})\citenamefont {Zhang}, \citenamefont {Cole},\
		and\ \citenamefont {Das~Sarma}}]{PhysRevLett.122.187001}%
	\BibitemOpen
	\bibfield  {author} {\bibinfo {author} {\bibfnamefont {R.~X.}\ \bibnamefont
			{Zhang}}, \bibinfo {author} {\bibfnamefont {W.~S.}\ \bibnamefont {Cole}}, \
		and\ \bibinfo {author} {\bibfnamefont {S.}~\bibnamefont {Das~Sarma}},\
	}\bibfield  {title} {\enquote {\bibinfo {title} {Helical hinge {M}ajorana
				modes in iron-based superconductors},}\ }\href {\doibase
		10.1103/PhysRevLett.122.187001} {\bibfield  {journal} {\bibinfo  {journal}
			{Phys. Rev. Lett.}\ }\textbf {\bibinfo {volume} {122}},\ \bibinfo {pages}
		{187001} (\bibinfo {year} {2019}{\natexlab{b}})}\BibitemShut {NoStop}%
	\bibitem [{\citenamefont {Pan}\ \emph {et~al.}(2019)\citenamefont {Pan},
		\citenamefont {Yang}, \citenamefont {Chen}, \citenamefont {Xu}, \citenamefont
		{Liu},\ and\ \citenamefont {Liu}}]{PhysRevLett.123.156801}%
	\BibitemOpen
	\bibfield  {author} {\bibinfo {author} {\bibfnamefont {X.~H.}\ \bibnamefont
			{Pan}}, \bibinfo {author} {\bibfnamefont {K.~J.}\ \bibnamefont {Yang}},
		\bibinfo {author} {\bibfnamefont {L.}~\bibnamefont {Chen}}, \bibinfo {author}
		{\bibfnamefont {G.}~\bibnamefont {Xu}}, \bibinfo {author} {\bibfnamefont
			{C.~X.}\ \bibnamefont {Liu}}, \ and\ \bibinfo {author} {\bibfnamefont
			{X.}~\bibnamefont {Liu}},\ }\bibfield  {title} {\enquote {\bibinfo {title}
			{Lattice-symmetry-assisted second-order topological superconductors and
				{M}ajorana patterns},}\ }\href {\doibase 10.1103/PhysRevLett.123.156801}
	{\bibfield  {journal} {\bibinfo  {journal} {Phys. Rev. Lett.}\ }\textbf
		{\bibinfo {volume} {123}},\ \bibinfo {pages} {156801} (\bibinfo {year}
		{2019})}\BibitemShut {NoStop}%
	\bibitem [{\citenamefont {Yang}\ \emph {et~al.}(2021)\citenamefont {Yang},
		\citenamefont {Li}, \citenamefont {Duan},\ and\ \citenamefont
		{Xu}}]{PhysRevB.103.085408}%
	\BibitemOpen
	\bibfield  {author} {\bibinfo {author} {\bibfnamefont {Y.~B.}\ \bibnamefont
			{Yang}}, \bibinfo {author} {\bibfnamefont {K.}~\bibnamefont {Li}}, \bibinfo
		{author} {\bibfnamefont {L.~M.}\ \bibnamefont {Duan}}, \ and\ \bibinfo
		{author} {\bibfnamefont {Y.}~\bibnamefont {Xu}},\ }\bibfield  {title}
	{\enquote {\bibinfo {title} {Higher-order topological anderson insulators},}\
	}\href {\doibase 10.1103/PhysRevB.103.085408} {\bibfield  {journal} {\bibinfo
			{journal} {Phys. Rev. B}\ }\textbf {\bibinfo {volume} {103}},\ \bibinfo
		{pages} {085408} (\bibinfo {year} {2021})}\BibitemShut {NoStop}%
	\bibitem [{\citenamefont {Liu}\ \emph {et~al.}(2021)\citenamefont {Liu},
		\citenamefont {Hu}, \citenamefont {Chen}, \citenamefont {Zhou},\ and\
		\citenamefont {Xu}}]{PhysRevB.103.L201115}%
	\BibitemOpen
	\bibfield  {author} {\bibinfo {author} {\bibfnamefont {Z.~R.}\ \bibnamefont
			{Liu}}, \bibinfo {author} {\bibfnamefont {L.~H.}\ \bibnamefont {Hu}},
		\bibinfo {author} {\bibfnamefont {C.~Z.}\ \bibnamefont {Chen}}, \bibinfo
		{author} {\bibfnamefont {B.}~\bibnamefont {Zhou}}, \ and\ \bibinfo {author}
		{\bibfnamefont {D.~H.}\ \bibnamefont {Xu}},\ }\bibfield  {title} {\enquote
		{\bibinfo {title} {Topological excitonic corner states and nodal phase in
				bilayer quantum spin hall insulators},}\ }\href {\doibase
		10.1103/PhysRevB.103.L201115} {\bibfield  {journal} {\bibinfo  {journal}
			{Phys. Rev. B}\ }\textbf {\bibinfo {volume} {103}},\ \bibinfo {pages}
		{L201115} (\bibinfo {year} {2021})}\BibitemShut {NoStop}%
	\bibitem [{\citenamefont {Wang}\ \emph {et~al.}(2020)\citenamefont {Wang},
		\citenamefont {Lin}, \citenamefont {Jiang}, \citenamefont {Guo},\ and\
		\citenamefont {Jiang}}]{PhysRevLett.125.146401}%
	\BibitemOpen
	\bibfield  {author} {\bibinfo {author} {\bibfnamefont {H.~X.}\ \bibnamefont
			{Wang}}, \bibinfo {author} {\bibfnamefont {Z.~K.}\ \bibnamefont {Lin}},
		\bibinfo {author} {\bibfnamefont {B.}~\bibnamefont {Jiang}}, \bibinfo
		{author} {\bibfnamefont {G.~Y.}\ \bibnamefont {Guo}}, \ and\ \bibinfo
		{author} {\bibfnamefont {J.~H.}\ \bibnamefont {Jiang}},\ }\bibfield  {title}
	{\enquote {\bibinfo {title} {Higher-order {W}eyl semimetals},}\ }\href
	{\doibase 10.1103/PhysRevLett.125.146401} {\bibfield  {journal} {\bibinfo
			{journal} {Phys. Rev. Lett.}\ }\textbf {\bibinfo {volume} {125}},\ \bibinfo
		{pages} {146401} (\bibinfo {year} {2020})}\BibitemShut {NoStop}%
	\bibitem [{\citenamefont {Lu}\ \emph {et~al.}(2023{\natexlab{a}})\citenamefont
		{Lu}, \citenamefont {Zhang}, \citenamefont {Wang}, \citenamefont {Ai},\ and\
		\citenamefont {Liu}}]{PhysRevB.107.125118}%
	\BibitemOpen
	\bibfield  {author} {\bibinfo {author} {\bibfnamefont {C.}~\bibnamefont
			{Lu}}, \bibinfo {author} {\bibfnamefont {M.}~\bibnamefont {Zhang}}, \bibinfo
		{author} {\bibfnamefont {H.}~\bibnamefont {Wang}}, \bibinfo {author}
		{\bibfnamefont {Q.}~\bibnamefont {Ai}}, \ and\ \bibinfo {author}
		{\bibfnamefont {T.}~\bibnamefont {Liu}},\ }\bibfield  {title} {\enquote
		{\bibinfo {title} {Topological quantum transition driven by charge-phonon
				coupling in higher-order topological insulators},}\ }\href {\doibase
		10.1103/PhysRevB.107.125118} {\bibfield  {journal} {\bibinfo  {journal}
			{Phys. Rev. B}\ }\textbf {\bibinfo {volume} {107}},\ \bibinfo {pages}
		{125118} (\bibinfo {year} {2023}{\natexlab{a}})}\BibitemShut {NoStop}%
	\bibitem [{\citenamefont {Lu}\ \emph {et~al.}(2023{\natexlab{b}})\citenamefont
		{Lu}, \citenamefont {Cai}, \citenamefont {Zhang}, \citenamefont {Wang},
		\citenamefont {Ai},\ and\ \citenamefont {Liu}}]{PhysRevB.107.165403}%
	\BibitemOpen
	\bibfield  {author} {\bibinfo {author} {\bibfnamefont {C.}~\bibnamefont
			{Lu}}, \bibinfo {author} {\bibfnamefont {Z.-F.}\ \bibnamefont {Cai}},
		\bibinfo {author} {\bibfnamefont {M.}~\bibnamefont {Zhang}}, \bibinfo
		{author} {\bibfnamefont {H.}~\bibnamefont {Wang}}, \bibinfo {author}
		{\bibfnamefont {Q.}~\bibnamefont {Ai}}, \ and\ \bibinfo {author}
		{\bibfnamefont {T.}~\bibnamefont {Liu}},\ }\bibfield  {title} {\enquote
		{\bibinfo {title} {Effects of disorder on {T}houless pumping in higher-order
				topological insulators},}\ }\href {\doibase 10.1103/PhysRevB.107.165403}
	{\bibfield  {journal} {\bibinfo  {journal} {Phys. Rev. B}\ }\textbf {\bibinfo
			{volume} {107}},\ \bibinfo {pages} {165403} (\bibinfo {year}
		{2023}{\natexlab{b}})}\BibitemShut {NoStop}%
	\bibitem [{\citenamefont {Kang}\ \emph {et~al.}(2023)\citenamefont {Kang},
		\citenamefont {Liu}, \citenamefont {Yan}, \citenamefont {Yang}, \citenamefont
		{Huang}, \citenamefont {Wei}, \citenamefont {Qiu}, \citenamefont {Dong},
		\citenamefont {Yang},\ and\ \citenamefont {Nori}}]{Kang2023}%
	\BibitemOpen
	\bibfield  {author} {\bibinfo {author} {\bibfnamefont {J.}~\bibnamefont
			{Kang}}, \bibinfo {author} {\bibfnamefont {T.}~\bibnamefont {Liu}}, \bibinfo
		{author} {\bibfnamefont {M.}~\bibnamefont {Yan}}, \bibinfo {author}
		{\bibfnamefont {D.}~\bibnamefont {Yang}}, \bibinfo {author} {\bibfnamefont
			{X.}~\bibnamefont {Huang}}, \bibinfo {author} {\bibfnamefont
			{R.}~\bibnamefont {Wei}}, \bibinfo {author} {\bibfnamefont {J.}~\bibnamefont
			{Qiu}}, \bibinfo {author} {\bibfnamefont {G.}~\bibnamefont {Dong}}, \bibinfo
		{author} {\bibfnamefont {Z.}~\bibnamefont {Yang}}, \ and\ \bibinfo {author}
		{\bibfnamefont {F.}~\bibnamefont {Nori}},\ }\bibfield  {title} {\enquote
		{\bibinfo {title} {Observation of square‐root higher-order topological
				states in photonic waveguide arrays},}\ }\href {\doibase
		10.1002/lpor.202200499} {\bibfield  {journal} {\bibinfo  {journal} {Laser
				Photonics Rev.}\ }\textbf {\bibinfo {volume} {17}},\ \bibinfo {pages}
		{2200499} (\bibinfo {year} {2023})}\BibitemShut {NoStop}%
	\bibitem [{\citenamefont {Yang}\ \emph {et~al.}(2024)\citenamefont {Yang},
		\citenamefont {Wang}, \citenamefont {Li},\ and\ \citenamefont
		{Xu}}]{Yang2024}%
	\BibitemOpen
	\bibfield  {author} {\bibinfo {author} {\bibfnamefont {Y.-B.}\ \bibnamefont
			{Yang}}, \bibinfo {author} {\bibfnamefont {J.-H.}\ \bibnamefont {Wang}},
		\bibinfo {author} {\bibfnamefont {K.}~\bibnamefont {Li}}, \ and\ \bibinfo
		{author} {\bibfnamefont {Y.}~\bibnamefont {Xu}},\ }\bibfield  {title}
	{\enquote {\bibinfo {title} {Higher-order topological phases in crystalline
				and non-crystalline systems: a review},}\ }\href {\doibase
		10.1088/1361-648x/ad3abd} {\bibfield  {journal} {\bibinfo  {journal} {J.
				Phys. Condens. Matter}\ }\textbf {\bibinfo {volume} {36}},\ \bibinfo {pages}
		{283002} (\bibinfo {year} {2024})}\BibitemShut {NoStop}%
	\bibitem [{\citenamefont {Wang}\ \emph {et~al.}(2021)\citenamefont {Wang},
		\citenamefont {Yang}, \citenamefont {Dai},\ and\ \citenamefont
		{Xu}}]{PhysRevLett.126.206404}%
	\BibitemOpen
	\bibfield  {author} {\bibinfo {author} {\bibfnamefont {J.-H.}\ \bibnamefont
			{Wang}}, \bibinfo {author} {\bibfnamefont {Y.-B.}\ \bibnamefont {Yang}},
		\bibinfo {author} {\bibfnamefont {N.}~\bibnamefont {Dai}}, \ and\ \bibinfo
		{author} {\bibfnamefont {Y.}~\bibnamefont {Xu}},\ }\bibfield  {title}
	{\enquote {\bibinfo {title} {Structural-disorder-induced second-order
				topological insulators in three dimensions},}\ }\href {\doibase
		10.1103/PhysRevLett.126.206404} {\bibfield  {journal} {\bibinfo  {journal}
			{Phys. Rev. Lett.}\ }\textbf {\bibinfo {volume} {126}},\ \bibinfo {pages}
		{206404} (\bibinfo {year} {2021})}\BibitemShut {NoStop}%
	\bibitem [{\citenamefont {Agarwala}\ \emph {et~al.}(2020)\citenamefont
		{Agarwala}, \citenamefont {Juri\ifmmode \check{c}\else
			\v{c}\fi{}i\ifmmode~\acute{c}\else \'{c}\fi{}},\ and\ \citenamefont
		{Roy}}]{PhysRevResearch.2.012067}%
	\BibitemOpen
	\bibfield  {author} {\bibinfo {author} {\bibfnamefont {A.}~\bibnamefont
			{Agarwala}}, \bibinfo {author} {\bibfnamefont {V.}~\bibnamefont {Juri\ifmmode
				\check{c}\else \v{c}\fi{}i\ifmmode~\acute{c}\else \'{c}\fi{}}}, \ and\
		\bibinfo {author} {\bibfnamefont {B.}~\bibnamefont {Roy}},\ }\bibfield
	{title} {\enquote {\bibinfo {title} {Higher-order topological insulators in
				amorphous solids},}\ }\href {\doibase 10.1103/PhysRevResearch.2.012067}
	{\bibfield  {journal} {\bibinfo  {journal} {Phys. Rev. Res.}\ }\textbf
		{\bibinfo {volume} {2}},\ \bibinfo {pages} {012067} (\bibinfo {year}
		{2020})}\BibitemShut {NoStop}%
	\bibitem [{\citenamefont {Manna}\ \emph {et~al.}(2022)\citenamefont {Manna},
		\citenamefont {Nandy},\ and\ \citenamefont {Roy}}]{PhysRevB.105.L201301}%
	\BibitemOpen
	\bibfield  {author} {\bibinfo {author} {\bibfnamefont {S.}~\bibnamefont
			{Manna}}, \bibinfo {author} {\bibfnamefont {S.}~\bibnamefont {Nandy}}, \ and\
		\bibinfo {author} {\bibfnamefont {B.}~\bibnamefont {Roy}},\ }\bibfield
	{title} {\enquote {\bibinfo {title} {Higher-order topological phases on
				fractal lattices},}\ }\href {\doibase 10.1103/PhysRevB.105.L201301}
	{\bibfield  {journal} {\bibinfo  {journal} {Phys. Rev. B}\ }\textbf {\bibinfo
			{volume} {105}},\ \bibinfo {pages} {L201301} (\bibinfo {year}
		{2022})}\BibitemShut {NoStop}%
	\bibitem [{\citenamefont {Rudner}\ and\ \citenamefont
		{Lindner}(2020)}]{Rudner2020}%
	\BibitemOpen
	\bibfield  {author} {\bibinfo {author} {\bibfnamefont {M.~S.}\ \bibnamefont
			{Rudner}}\ and\ \bibinfo {author} {\bibfnamefont {N.~H.}\ \bibnamefont
			{Lindner}},\ }\bibfield  {title} {\enquote {\bibinfo {title} {Band structure
				engineering and non-equilibrium dynamics in {F}loquet topological
				insulators},}\ }\href {\doibase 10.1038/s42254-020-0170-z} {\bibfield
		{journal} {\bibinfo  {journal} {Nat. Rev. Phys.}\ }\textbf {\bibinfo {volume}
			{2}},\ \bibinfo {pages} {229} (\bibinfo {year} {2020})}\BibitemShut {NoStop}%
	\bibitem [{\citenamefont {Lababidi}\ \emph {et~al.}(2014)\citenamefont
		{Lababidi}, \citenamefont {Satija},\ and\ \citenamefont
		{Zhao}}]{PhysRevLett.112.026805}%
	\BibitemOpen
	\bibfield  {author} {\bibinfo {author} {\bibfnamefont {Mahmoud}\ \bibnamefont
			{Lababidi}}, \bibinfo {author} {\bibfnamefont {Indubala~I.}\ \bibnamefont
			{Satija}}, \ and\ \bibinfo {author} {\bibfnamefont {Erhai}\ \bibnamefont
			{Zhao}},\ }\bibfield  {title} {\enquote {\bibinfo {title}
			{Counter-propagating edge modes and topological phases of a kicked quantum
				hall system},}\ }\href {\doibase 10.1103/PhysRevLett.112.026805} {\bibfield
		{journal} {\bibinfo  {journal} {Phys. Rev. Lett.}\ }\textbf {\bibinfo
			{volume} {112}},\ \bibinfo {pages} {026805} (\bibinfo {year}
		{2014})}\BibitemShut {NoStop}%
	\bibitem [{\citenamefont {Lindner}\ \emph {et~al.}(2011)\citenamefont
		{Lindner}, \citenamefont {Refael},\ and\ \citenamefont
		{Galitski}}]{Lindner2011}%
	\BibitemOpen
	\bibfield  {author} {\bibinfo {author} {\bibfnamefont {N.~H.}\ \bibnamefont
			{Lindner}}, \bibinfo {author} {\bibfnamefont {G.}~\bibnamefont {Refael}}, \
		and\ \bibinfo {author} {\bibfnamefont {V.}~\bibnamefont {Galitski}},\
	}\bibfield  {title} {\enquote {\bibinfo {title} {Floquet topological
				insulator in semiconductor quantum wells},}\ }\href {\doibase
		10.1038/nphys1926} {\bibfield  {journal} {\bibinfo  {journal} {Nat. Phys.}\
		}\textbf {\bibinfo {volume} {7}},\ \bibinfo {pages} {490} (\bibinfo {year}
		{2011})}\BibitemShut {NoStop}%
	\bibitem [{\citenamefont {Maczewsky}\ \emph {et~al.}(2020)\citenamefont
		{Maczewsky}, \citenamefont {H\"{o}ckendorf}, \citenamefont {Kremer},
		\citenamefont {Biesenthal}, \citenamefont {Heinrich}, \citenamefont
		{Alvermann}, \citenamefont {Fehske},\ and\ \citenamefont
		{Szameit}}]{Maczewsky2020}%
	\BibitemOpen
	\bibfield  {author} {\bibinfo {author} {\bibfnamefont {L.~J.}\ \bibnamefont
			{Maczewsky}}, \bibinfo {author} {\bibfnamefont {B.}~\bibnamefont
			{H\"{o}ckendorf}}, \bibinfo {author} {\bibfnamefont {M.}~\bibnamefont
			{Kremer}}, \bibinfo {author} {\bibfnamefont {T.}~\bibnamefont {Biesenthal}},
		\bibinfo {author} {\bibfnamefont {M.}~\bibnamefont {Heinrich}}, \bibinfo
		{author} {\bibfnamefont {A.}~\bibnamefont {Alvermann}}, \bibinfo {author}
		{\bibfnamefont {H.}~\bibnamefont {Fehske}}, \ and\ \bibinfo {author}
		{\bibfnamefont {A.}~\bibnamefont {Szameit}},\ }\bibfield  {title} {\enquote
		{\bibinfo {title} {Fermionic time-reversal symmetry in a photonic topological
				insulator},}\ }\href {\doibase 10.1038/s41563-020-0641-8} {\bibfield
		{journal} {\bibinfo  {journal} {Nat. Mater.}\ }\textbf {\bibinfo {volume}
			{19}},\ \bibinfo {pages} {855} (\bibinfo {year} {2020})}\BibitemShut
	{NoStop}%
	\bibitem [{\citenamefont {Pyrialakos}\ \emph {et~al.}(2022)\citenamefont
		{Pyrialakos}, \citenamefont {Beck}, \citenamefont {Heinrich}, \citenamefont
		{Maczewsky}, \citenamefont {Kantartzis}, \citenamefont {Khajavikhan},
		\citenamefont {Szameit},\ and\ \citenamefont
		{Christodoulides}}]{Pyrialakos2022}%
	\BibitemOpen
	\bibfield  {author} {\bibinfo {author} {\bibfnamefont {G.~G.}\ \bibnamefont
			{Pyrialakos}}, \bibinfo {author} {\bibfnamefont {J.}~\bibnamefont {Beck}},
		\bibinfo {author} {\bibfnamefont {M.}~\bibnamefont {Heinrich}}, \bibinfo
		{author} {\bibfnamefont {L.~J.}\ \bibnamefont {Maczewsky}}, \bibinfo {author}
		{\bibfnamefont {N.~V.}\ \bibnamefont {Kantartzis}}, \bibinfo {author}
		{\bibfnamefont {M.}~\bibnamefont {Khajavikhan}}, \bibinfo {author}
		{\bibfnamefont {A.}~\bibnamefont {Szameit}}, \ and\ \bibinfo {author}
		{\bibfnamefont {D.~N.}\ \bibnamefont {Christodoulides}},\ }\bibfield  {title}
	{\enquote {\bibinfo {title} {Bimorphic {F}loquet topological insulators},}\
	}\href {\doibase 10.1038/s41563-022-01238-w} {\bibfield  {journal} {\bibinfo
			{journal} {Nat. Mater.}\ }\textbf {\bibinfo {volume} {21}},\ \bibinfo {pages}
		{634} (\bibinfo {year} {2022})}\BibitemShut {NoStop}%
	\bibitem [{\citenamefont {Rudner}\ \emph {et~al.}(2013)\citenamefont {Rudner},
		\citenamefont {Lindner}, \citenamefont {Berg},\ and\ \citenamefont
		{Levin}}]{rudner2013anomalous}%
	\BibitemOpen
	\bibfield  {author} {\bibinfo {author} {\bibfnamefont {M.~S.}\ \bibnamefont
			{Rudner}}, \bibinfo {author} {\bibfnamefont {N.~H.}\ \bibnamefont {Lindner}},
		\bibinfo {author} {\bibfnamefont {E.}~\bibnamefont {Berg}}, \ and\ \bibinfo
		{author} {\bibfnamefont {M.}~\bibnamefont {Levin}},\ }\bibfield  {title}
	{\enquote {\bibinfo {title} {Anomalous edge states and the bulk-edge
				correspondence for periodically driven two-dimensional systems},}\ }\href
	{http://dx.doi.org/10.1103/PhysRevX.3.031005} {\bibfield  {journal} {\bibinfo
			{journal} {Phys. Rev. X}\ }\textbf {\bibinfo {volume} {3}},\ \bibinfo
		{pages} {031005} (\bibinfo {year} {2013})}\BibitemShut {NoStop}%
	\bibitem [{\citenamefont {Peng}\ \emph {et~al.}(2016)\citenamefont {Peng},
		\citenamefont {Qin}, \citenamefont {Zhao}, \citenamefont {Shen},
		\citenamefont {Xu}, \citenamefont {Jia},\ and\ \citenamefont
		{Zhu}}]{Peng2016}%
	\BibitemOpen
	\bibfield  {author} {\bibinfo {author} {\bibfnamefont {Y.-G.}\ \bibnamefont
			{Peng}}, \bibinfo {author} {\bibfnamefont {C.-Z.}\ \bibnamefont {Qin}},
		\bibinfo {author} {\bibfnamefont {D.-G.}\ \bibnamefont {Zhao}}, \bibinfo
		{author} {\bibfnamefont {Y.-X.}\ \bibnamefont {Shen}}, \bibinfo {author}
		{\bibfnamefont {M.}~\bibnamefont {Xu}, \bibfnamefont {X.-Y.and~Bao}},
		\bibinfo {author} {\bibfnamefont {H.}~\bibnamefont {Jia}}, \ and\ \bibinfo
		{author} {\bibfnamefont {X.-F.}\ \bibnamefont {Zhu}},\ }\bibfield  {title}
	{\enquote {\bibinfo {title} {Experimental demonstration of anomalous
				{F}loquet topological insulator for sound},}\ }\href {\doibase
		10.1038/ncomms13368} {\bibfield  {journal} {\bibinfo  {journal} {Nat.
				Commun.}\ }\textbf {\bibinfo {volume} {7}},\ \bibinfo {pages} {13368}
		(\bibinfo {year} {2016})}\BibitemShut {NoStop}%
	\bibitem [{\citenamefont {Maczewsky}\ \emph {et~al.}(2017)\citenamefont
		{Maczewsky}, \citenamefont {Zeuner}, \citenamefont {Nolte},\ and\
		\citenamefont {Szameit}}]{maczewsky2017observation}%
	\BibitemOpen
	\bibfield  {author} {\bibinfo {author} {\bibfnamefont {L.~J.}\ \bibnamefont
			{Maczewsky}}, \bibinfo {author} {\bibfnamefont {J.~M.}\ \bibnamefont
			{Zeuner}}, \bibinfo {author} {\bibfnamefont {S.}~\bibnamefont {Nolte}}, \
		and\ \bibinfo {author} {\bibfnamefont {A.}~\bibnamefont {Szameit}},\
	}\bibfield  {title} {\enquote {\bibinfo {title} {Observation of photonic
				anomalous {F}loquet topological insulators},}\ }\href
	{http://dx.doi.org/10.1038/ncomms13756} {\bibfield  {journal} {\bibinfo
			{journal} {Nat. Commun.}\ }\textbf {\bibinfo {volume} {8}},\ \bibinfo {pages}
		{13756} (\bibinfo {year} {2017})}\BibitemShut {NoStop}%
	\bibitem [{\citenamefont {Yao}\ \emph {et~al.}(2017)\citenamefont {Yao},
		\citenamefont {Yan},\ and\ \citenamefont {Wang}}]{PhysRevB.96.195303}%
	\BibitemOpen
	\bibfield  {author} {\bibinfo {author} {\bibfnamefont {S.}~\bibnamefont
			{Yao}}, \bibinfo {author} {\bibfnamefont {Z.}~\bibnamefont {Yan}}, \ and\
		\bibinfo {author} {\bibfnamefont {Z.}~\bibnamefont {Wang}},\ }\bibfield
	{title} {\enquote {\bibinfo {title} {Topological invariants of {F}loquet
				systems: {G}eneral formulation, special properties, and {F}loquet topological
				defects},}\ }\href {\doibase 10.1103/PhysRevB.96.195303} {\bibfield
		{journal} {\bibinfo  {journal} {Phys. Rev. B}\ }\textbf {\bibinfo {volume}
			{96}},\ \bibinfo {pages} {195303} (\bibinfo {year} {2017})}\BibitemShut
	{NoStop}%
	\bibitem [{\citenamefont {Zhou}\ and\ \citenamefont
		{Lee}(2019)}]{PhysRevB.99.235112}%
	\BibitemOpen
	\bibfield  {author} {\bibinfo {author} {\bibfnamefont {H.}~\bibnamefont
			{Zhou}}\ and\ \bibinfo {author} {\bibfnamefont {J.~Y.}\ \bibnamefont {Lee}},\
	}\bibfield  {title} {\enquote {\bibinfo {title} {Periodic table for
				topological bands with non-{H}ermitian symmetries},}\ }\href {\doibase
		10.1103/PhysRevB.99.235112} {\bibfield  {journal} {\bibinfo  {journal} {Phys.
				Rev. B}\ }\textbf {\bibinfo {volume} {99}},\ \bibinfo {pages} {235112}
		(\bibinfo {year} {2019})}\BibitemShut {NoStop}%
	\bibitem [{\citenamefont {Wintersperger}\ \emph {et~al.}(2020)\citenamefont
		{Wintersperger}, \citenamefont {Braun}, \citenamefont {\"{U}nal},
		\citenamefont {Eckardt}, \citenamefont {Liberto}, \citenamefont {Goldman},
		\citenamefont {Bloch},\ and\ \citenamefont
		{Aidelsburger}}]{Wintersperger2020}%
	\BibitemOpen
	\bibfield  {author} {\bibinfo {author} {\bibfnamefont {K.}~\bibnamefont
			{Wintersperger}}, \bibinfo {author} {\bibfnamefont {C.}~\bibnamefont
			{Braun}}, \bibinfo {author} {\bibfnamefont {F.~N.}\ \bibnamefont {\"{U}nal}},
		\bibinfo {author} {\bibfnamefont {A.}~\bibnamefont {Eckardt}}, \bibinfo
		{author} {\bibfnamefont {M.~D.}\ \bibnamefont {Liberto}}, \bibinfo {author}
		{\bibfnamefont {N.}~\bibnamefont {Goldman}}, \bibinfo {author} {\bibfnamefont
			{I.}~\bibnamefont {Bloch}}, \ and\ \bibinfo {author} {\bibfnamefont
			{M.}~\bibnamefont {Aidelsburger}},\ }\bibfield  {title} {\enquote {\bibinfo
			{title} {Realization of an anomalous {F}loquet topological system with
				ultracold atoms},}\ }\href {\doibase 10.1038/s41567-020-0949-y} {\bibfield
		{journal} {\bibinfo  {journal} {Nat. Phys.}\ }\textbf {\bibinfo {volume}
			{16}},\ \bibinfo {pages} {1058} (\bibinfo {year} {2020})}\BibitemShut
	{NoStop}%
	\bibitem [{\citenamefont {Ladovrechis}\ and\ \citenamefont
		{Fulga}(2019)}]{PhysRevB.99.195426}%
	\BibitemOpen
	\bibfield  {author} {\bibinfo {author} {\bibfnamefont {K.}~\bibnamefont
			{Ladovrechis}}\ and\ \bibinfo {author} {\bibfnamefont {I.~C.}\ \bibnamefont
			{Fulga}},\ }\bibfield  {title} {\enquote {\bibinfo {title} {Anomalous
				{F}loquet topological crystalline insulators},}\ }\href {\doibase
		10.1103/PhysRevB.99.195426} {\bibfield  {journal} {\bibinfo  {journal} {Phys.
				Rev. B}\ }\textbf {\bibinfo {volume} {99}},\ \bibinfo {pages} {195426}
		(\bibinfo {year} {2019})}\BibitemShut {NoStop}%
	\bibitem [{\citenamefont {Peng}\ and\ \citenamefont
		{Refael}(2019)}]{PhysRevLett.123.016806}%
	\BibitemOpen
	\bibfield  {author} {\bibinfo {author} {\bibfnamefont {Y.}~\bibnamefont
			{Peng}}\ and\ \bibinfo {author} {\bibfnamefont {G.}~\bibnamefont {Refael}},\
	}\bibfield  {title} {\enquote {\bibinfo {title} {Floquet second-order
				topological insulators from nonsymmorphic space-time symmetries},}\ }\href
	{\doibase 10.1103/PhysRevLett.123.016806} {\bibfield  {journal} {\bibinfo
			{journal} {Phys. Rev. Lett.}\ }\textbf {\bibinfo {volume} {123}},\ \bibinfo
		{pages} {016806} (\bibinfo {year} {2019})}\BibitemShut {NoStop}%
	\bibitem [{\citenamefont {Hu}\ \emph {et~al.}(2020)\citenamefont {Hu},
		\citenamefont {Huang}, \citenamefont {Zhao},\ and\ \citenamefont
		{Liu}}]{hu2020dynamical}%
	\BibitemOpen
	\bibfield  {author} {\bibinfo {author} {\bibfnamefont {H.}~\bibnamefont
			{Hu}}, \bibinfo {author} {\bibfnamefont {B.}~\bibnamefont {Huang}}, \bibinfo
		{author} {\bibfnamefont {E.}~\bibnamefont {Zhao}}, \ and\ \bibinfo {author}
		{\bibfnamefont {W.~V.}\ \bibnamefont {Liu}},\ }\bibfield  {title} {\enquote
		{\bibinfo {title} {Dynamical singularities of {F}loquet higher-order
				topological insulators},}\ }\href {\doibase 10.1103/PhysRevLett.124.057001}
	{\bibfield  {journal} {\bibinfo  {journal} {Phys. Rev. Lett.}\ }\textbf
		{\bibinfo {volume} {124}},\ \bibinfo {pages} {057001} (\bibinfo {year}
		{2020})}\BibitemShut {NoStop}%
	\bibitem [{\citenamefont {Huang}\ and\ \citenamefont
		{Liu}(2020)}]{huang2020floquet}%
	\BibitemOpen
	\bibfield  {author} {\bibinfo {author} {\bibfnamefont {B.}~\bibnamefont
			{Huang}}\ and\ \bibinfo {author} {\bibfnamefont {W.~V.}\ \bibnamefont
			{Liu}},\ }\bibfield  {title} {\enquote {\bibinfo {title} {Floquet
				higher-order topological insulators with anomalous dynamical polarization},}\
	}\href {\doibase 10.1103/PhysRevLett.124.216601} {\bibfield  {journal}
		{\bibinfo  {journal} {Phys. Rev. Lett.}\ }\textbf {\bibinfo {volume} {124}},\
		\bibinfo {pages} {216601} (\bibinfo {year} {2020})}\BibitemShut {NoStop}%
	\bibitem [{\citenamefont {Yu}\ \emph {et~al.}(2021)\citenamefont {Yu},
		\citenamefont {Zhang},\ and\ \citenamefont {Song}}]{Yu2021}%
	\BibitemOpen
	\bibfield  {author} {\bibinfo {author} {\bibfnamefont {J.}~\bibnamefont
			{Yu}}, \bibinfo {author} {\bibfnamefont {R.-X.}\ \bibnamefont {Zhang}}, \
		and\ \bibinfo {author} {\bibfnamefont {Z.-D.}\ \bibnamefont {Song}},\
	}\bibfield  {title} {\enquote {\bibinfo {title} {Dynamical symmetry
				indicators for {F}loquet crystals},}\ }\href {\doibase
		10.1038/s41467-021-26092-3} {\bibfield  {journal} {\bibinfo  {journal} {Nat.
				Commun.}\ }\textbf {\bibinfo {volume} {12}},\ \bibinfo {pages} {5985}
		(\bibinfo {year} {2021})}\BibitemShut {NoStop}%
	\bibitem [{\citenamefont {Nag}\ \emph {et~al.}(2021)\citenamefont {Nag},
		\citenamefont {Juri\ifmmode \check{c}\else \v{c}\fi{}i\ifmmode~\acute{c}\else
			\'{c}\fi{}},\ and\ \citenamefont {Roy}}]{PhysRevB.103.115308}%
	\BibitemOpen
	\bibfield  {author} {\bibinfo {author} {\bibfnamefont {T.}~\bibnamefont
			{Nag}}, \bibinfo {author} {\bibfnamefont {V.}~\bibnamefont {Juri\ifmmode
				\check{c}\else \v{c}\fi{}i\ifmmode~\acute{c}\else \'{c}\fi{}}}, \ and\
		\bibinfo {author} {\bibfnamefont {B.}~\bibnamefont {Roy}},\ }\bibfield
	{title} {\enquote {\bibinfo {title} {Hierarchy of higher-order {F}loquet
				topological phases in three dimensions},}\ }\href {\doibase
		10.1103/PhysRevB.103.115308} {\bibfield  {journal} {\bibinfo  {journal}
			{Phys. Rev. B}\ }\textbf {\bibinfo {volume} {103}},\ \bibinfo {pages}
		{115308} (\bibinfo {year} {2021})}\BibitemShut {NoStop}%
	\bibitem [{\citenamefont {Ghosh}\ \emph {et~al.}(2020)\citenamefont {Ghosh},
		\citenamefont {Paul},\ and\ \citenamefont {Saha}}]{PhysRevB.101.235403}%
	\BibitemOpen
	\bibfield  {author} {\bibinfo {author} {\bibfnamefont {A.~K.}\ \bibnamefont
			{Ghosh}}, \bibinfo {author} {\bibfnamefont {G.~C.}\ \bibnamefont {Paul}}, \
		and\ \bibinfo {author} {\bibfnamefont {A.}~\bibnamefont {Saha}},\ }\bibfield
	{title} {\enquote {\bibinfo {title} {Higher order topological insulator via
				periodic driving},}\ }\href {\doibase 10.1103/PhysRevB.101.235403} {\bibfield
		{journal} {\bibinfo  {journal} {Phys. Rev. B}\ }\textbf {\bibinfo {volume}
			{101}},\ \bibinfo {pages} {235403} (\bibinfo {year} {2020})}\BibitemShut
	{NoStop}%
	\bibitem [{\citenamefont {Chaudhary}\ \emph {et~al.}(2020)\citenamefont
		{Chaudhary}, \citenamefont {Haim}, \citenamefont {Peng},\ and\ \citenamefont
		{Refael}}]{PhysRevResearch.2.043431}%
	\BibitemOpen
	\bibfield  {author} {\bibinfo {author} {\bibfnamefont {S.}~\bibnamefont
			{Chaudhary}}, \bibinfo {author} {\bibfnamefont {A.}~\bibnamefont {Haim}},
		\bibinfo {author} {\bibfnamefont {Y.}~\bibnamefont {Peng}}, \ and\ \bibinfo
		{author} {\bibfnamefont {G.}~\bibnamefont {Refael}},\ }\bibfield  {title}
	{\enquote {\bibinfo {title} {Phonon-induced {F}loquet topological phases
				protected by space-time symmetries},}\ }\href {\doibase
		10.1103/PhysRevResearch.2.043431} {\bibfield  {journal} {\bibinfo  {journal}
			{Phys. Rev. Res.}\ }\textbf {\bibinfo {volume} {2}},\ \bibinfo {pages}
		{043431} (\bibinfo {year} {2020})}\BibitemShut {NoStop}%
	\bibitem [{\citenamefont {Zhu}\ \emph {et~al.}(2022)\citenamefont {Zhu},
		\citenamefont {Xue}, \citenamefont {Gong}, \citenamefont {Chong},\ and\
		\citenamefont {Zhang}}]{zhu2022time}%
	\BibitemOpen
	\bibfield  {author} {\bibinfo {author} {\bibfnamefont {W.}~\bibnamefont
			{Zhu}}, \bibinfo {author} {\bibfnamefont {H.}~\bibnamefont {Xue}}, \bibinfo
		{author} {\bibfnamefont {J.}~\bibnamefont {Gong}}, \bibinfo {author}
		{\bibfnamefont {Y.}~\bibnamefont {Chong}}, \ and\ \bibinfo {author}
		{\bibfnamefont {B.}~\bibnamefont {Zhang}},\ }\bibfield  {title} {\enquote
		{\bibinfo {title} {Time-periodic corner states from {F}loquet higher-order
				topology},}\ }\href {http://dx.doi.org/10.1038/s41467-021-27552-6} {\bibfield
		{journal} {\bibinfo  {journal} {Nat. Commun.}\ }\textbf {\bibinfo {volume}
			{13}},\ \bibinfo {pages} {11} (\bibinfo {year} {2022})}\BibitemShut {NoStop}%
	\bibitem [{\citenamefont {Ghosh}\ and\ \citenamefont
		{Nag}(2024)}]{PhysRevB.110.125427}%
	\BibitemOpen
	\bibfield  {author} {\bibinfo {author} {\bibfnamefont {A.~K.}\ \bibnamefont
			{Ghosh}}\ and\ \bibinfo {author} {\bibfnamefont {A.}~\bibnamefont {Nag},
			\bibfnamefont {T.and~Saha}},\ }\bibfield  {title} {\enquote {\bibinfo {title}
			{Floquet second-order topological {A}nderson insulator hosting corner
				localized modes},}\ }\href {\doibase 10.1103/PhysRevB.110.125427} {\bibfield
		{journal} {\bibinfo  {journal} {Phys. Rev. B}\ }\textbf {\bibinfo {volume}
			{110}},\ \bibinfo {pages} {125427} (\bibinfo {year} {2024})}\BibitemShut
	{NoStop}%
	\bibitem [{\citenamefont {Hohenadler}\ and\ \citenamefont
		{Assaad}(2013)}]{Hohenadler2013}%
	\BibitemOpen
	\bibfield  {author} {\bibinfo {author} {\bibfnamefont {M}~\bibnamefont
			{Hohenadler}}\ and\ \bibinfo {author} {\bibfnamefont {F~F}\ \bibnamefont
			{Assaad}},\ }\bibfield  {title} {\enquote {\bibinfo {title} {Correlation
				effects in two-dimensional topological insulators},}\ }\href {\doibase
		10.1088/0953-8984/25/14/143201} {\bibfield  {journal} {\bibinfo  {journal}
			{J. Phys. Cond. Matt.}\ }\textbf {\bibinfo {volume} {25}},\ \bibinfo {pages}
		{143201} (\bibinfo {year} {2013})}\BibitemShut {NoStop}%
	\bibitem [{\citenamefont {Witczak-Krempa}\ \emph {et~al.}(2014)\citenamefont
		{Witczak-Krempa}, \citenamefont {Chen}, \citenamefont {Kim},\ and\
		\citenamefont {Balents}}]{WitczakKrempa2014}%
	\BibitemOpen
	\bibfield  {author} {\bibinfo {author} {\bibfnamefont {W.}~\bibnamefont
			{Witczak-Krempa}}, \bibinfo {author} {\bibfnamefont {G.}~\bibnamefont
			{Chen}}, \bibinfo {author} {\bibfnamefont {Y.~B.}\ \bibnamefont {Kim}}, \
		and\ \bibinfo {author} {\bibfnamefont {L.}~\bibnamefont {Balents}},\
	}\bibfield  {title} {\enquote {\bibinfo {title} {Correlated quantum phenomena
				in the strong spin-orbit regime},}\ }\href {\doibase
		10.1146/annurev-conmatphys-020911-125138} {\bibfield  {journal} {\bibinfo
			{journal} {Ann. Rev. Cond. Matt. Phys.}\ }\textbf {\bibinfo {volume} {5}},\
		\bibinfo {pages} {57} (\bibinfo {year} {2014})}\BibitemShut {NoStop}%
	\bibitem [{\citenamefont {Alexandrov}\ \emph {et~al.}(2015)\citenamefont
		{Alexandrov}, \citenamefont {Coleman},\ and\ \citenamefont
		{Erten}}]{PhysRevLett.114.177202}%
	\BibitemOpen
	\bibfield  {author} {\bibinfo {author} {\bibfnamefont {V.}~\bibnamefont
			{Alexandrov}}, \bibinfo {author} {\bibfnamefont {P.}~\bibnamefont {Coleman}},
		\ and\ \bibinfo {author} {\bibfnamefont {O.}~\bibnamefont {Erten}},\
	}\bibfield  {title} {\enquote {\bibinfo {title} {Kondo breakdown in
				topological {K}ondo insulators},}\ }\href {\doibase
		10.1103/PhysRevLett.114.177202} {\bibfield  {journal} {\bibinfo  {journal}
			{Phys. Rev. Lett.}\ }\textbf {\bibinfo {volume} {114}},\ \bibinfo {pages}
		{177202} (\bibinfo {year} {2015})}\BibitemShut {NoStop}%
	\bibitem [{\citenamefont {You}\ \emph {et~al.}(2018)\citenamefont {You},
		\citenamefont {Devakul}, \citenamefont {Burnell},\ and\ \citenamefont
		{Neupert}}]{PhysRevB.98.235102}%
	\BibitemOpen
	\bibfield  {author} {\bibinfo {author} {\bibfnamefont {Y.}~\bibnamefont
			{You}}, \bibinfo {author} {\bibfnamefont {T.}~\bibnamefont {Devakul}},
		\bibinfo {author} {\bibfnamefont {F.~J.}\ \bibnamefont {Burnell}}, \ and\
		\bibinfo {author} {\bibfnamefont {T.}~\bibnamefont {Neupert}},\ }\bibfield
	{title} {\enquote {\bibinfo {title} {Higher-order symmetry-protected
				topological states for interacting bosons and fermions},}\ }\href {\doibase
		10.1103/PhysRevB.98.235102} {\bibfield  {journal} {\bibinfo  {journal} {Phys.
				Rev. B}\ }\textbf {\bibinfo {volume} {98}},\ \bibinfo {pages} {235102}
		(\bibinfo {year} {2018})}\BibitemShut {NoStop}%
	\bibitem [{\citenamefont {Dubinkin}\ and\ \citenamefont
		{Hughes}(2019)}]{PhysRevB.99.235132}%
	\BibitemOpen
	\bibfield  {author} {\bibinfo {author} {\bibfnamefont {O.}~\bibnamefont
			{Dubinkin}}\ and\ \bibinfo {author} {\bibfnamefont {Taylor~L.}\ \bibnamefont
			{Hughes}},\ }\bibfield  {title} {\enquote {\bibinfo {title} {Higher-order
				bosonic topological phases in spin models},}\ }\href {\doibase
		10.1103/PhysRevB.99.235132} {\bibfield  {journal} {\bibinfo  {journal} {Phys.
				Rev. B}\ }\textbf {\bibinfo {volume} {99}},\ \bibinfo {pages} {235132}
		(\bibinfo {year} {2019})}\BibitemShut {NoStop}%
	\bibitem [{\citenamefont {Rasmussen}\ and\ \citenamefont
		{Lu}(2020)}]{PhysRevB.101.085137}%
	\BibitemOpen
	\bibfield  {author} {\bibinfo {author} {\bibfnamefont {A.}~\bibnamefont
			{Rasmussen}}\ and\ \bibinfo {author} {\bibfnamefont {Y.-M.}\ \bibnamefont
			{Lu}},\ }\bibfield  {title} {\enquote {\bibinfo {title} {Classification and
				construction of higher-order symmetry-protected topological phases of
				interacting bosons},}\ }\href {\doibase 10.1103/PhysRevB.101.085137}
	{\bibfield  {journal} {\bibinfo  {journal} {Phys. Rev. B}\ }\textbf {\bibinfo
			{volume} {101}},\ \bibinfo {pages} {085137} (\bibinfo {year}
		{2020})}\BibitemShut {NoStop}%
	\bibitem [{\citenamefont {Peng}\ \emph {et~al.}(2020)\citenamefont {Peng},
		\citenamefont {He},\ and\ \citenamefont {Lu}}]{PhysRevB.102.045110}%
	\BibitemOpen
	\bibfield  {author} {\bibinfo {author} {\bibfnamefont {C.}~\bibnamefont
			{Peng}}, \bibinfo {author} {\bibfnamefont {R.~Q.}\ \bibnamefont {He}}, \ and\
		\bibinfo {author} {\bibfnamefont {Z.~Y.}\ \bibnamefont {Lu}},\ }\bibfield
	{title} {\enquote {\bibinfo {title} {Correlation effects in quadrupole
				insulators: A quantum monte carlo study},}\ }\href {\doibase
		10.1103/PhysRevB.102.045110} {\bibfield  {journal} {\bibinfo  {journal}
			{Phys. Rev. B}\ }\textbf {\bibinfo {volume} {102}},\ \bibinfo {pages}
		{045110} (\bibinfo {year} {2020})}\BibitemShut {NoStop}%
	\bibitem [{\citenamefont {Montorsi}\ \emph {et~al.}(2022)\citenamefont
		{Montorsi}, \citenamefont {Bhattacharya}, \citenamefont {Gonz\'alez-Cuadra},
		\citenamefont {Lewenstein}, \citenamefont {Palumbo},\ and\ \citenamefont
		{Barbiero}}]{PhysRevB.106.L241115}%
	\BibitemOpen
	\bibfield  {author} {\bibinfo {author} {\bibfnamefont {A.}~\bibnamefont
			{Montorsi}}, \bibinfo {author} {\bibfnamefont {U.}~\bibnamefont
			{Bhattacharya}}, \bibinfo {author} {\bibfnamefont {Daniel}\ \bibnamefont
			{Gonz\'alez-Cuadra}}, \bibinfo {author} {\bibfnamefont {M.}~\bibnamefont
			{Lewenstein}}, \bibinfo {author} {\bibfnamefont {G.}~\bibnamefont {Palumbo}},
		\ and\ \bibinfo {author} {\bibfnamefont {L.}~\bibnamefont {Barbiero}},\
	}\bibfield  {title} {\enquote {\bibinfo {title} {Interacting second-order
				topological insulators in one-dimensional fermions with correlated
				hopping},}\ }\href {\doibase 10.1103/PhysRevB.106.L241115} {\bibfield
		{journal} {\bibinfo  {journal} {Phys. Rev. B}\ }\textbf {\bibinfo {volume}
			{106}},\ \bibinfo {pages} {L241115} (\bibinfo {year} {2022})}\BibitemShut
	{NoStop}%
	\bibitem [{\citenamefont {Li}\ \emph {et~al.}(2022)\citenamefont {Li},
		\citenamefont {Kee},\ and\ \citenamefont {Kim}}]{PhysRevB.106.155116}%
	\BibitemOpen
	\bibfield  {author} {\bibinfo {author} {\bibfnamefont {H.}~\bibnamefont
			{Li}}, \bibinfo {author} {\bibfnamefont {H.-Y.}\ \bibnamefont {Kee}}, \ and\
		\bibinfo {author} {\bibfnamefont {Y.~B.}\ \bibnamefont {Kim}},\ }\bibfield
	{title} {\enquote {\bibinfo {title} {Green's function approach to interacting
				higher-order topological insulators},}\ }\href {\doibase
		10.1103/PhysRevB.106.155116} {\bibfield  {journal} {\bibinfo  {journal}
			{Phys. Rev. B}\ }\textbf {\bibinfo {volume} {106}},\ \bibinfo {pages}
		{155116} (\bibinfo {year} {2022})}\BibitemShut {NoStop}%
	\bibitem [{\citenamefont {Kudo}\ \emph {et~al.}(2019)\citenamefont {Kudo},
		\citenamefont {Yoshida},\ and\ \citenamefont
		{Hatsugai}}]{PhysRevLett.123.196402}%
	\BibitemOpen
	\bibfield  {author} {\bibinfo {author} {\bibfnamefont {K.}~\bibnamefont
			{Kudo}}, \bibinfo {author} {\bibfnamefont {T.}~\bibnamefont {Yoshida}}, \
		and\ \bibinfo {author} {\bibfnamefont {Y.}~\bibnamefont {Hatsugai}},\
	}\bibfield  {title} {\enquote {\bibinfo {title} {Higher-order topological
				{M}ott insulators},}\ }\href {\doibase 10.1103/PhysRevLett.123.196402}
	{\bibfield  {journal} {\bibinfo  {journal} {Phys. Rev. Lett.}\ }\textbf
		{\bibinfo {volume} {123}},\ \bibinfo {pages} {196402} (\bibinfo {year}
		{2019})}\BibitemShut {NoStop}%
	\bibitem [{\citenamefont {Zhao}\ \emph {et~al.}(2021)\citenamefont {Zhao},
		\citenamefont {Qiang}, \citenamefont {Lu},\ and\ \citenamefont
		{Xie}}]{PhysRevLett.127.176601}%
	\BibitemOpen
	\bibfield  {author} {\bibinfo {author} {\bibfnamefont {P.~L.}\ \bibnamefont
			{Zhao}}, \bibinfo {author} {\bibfnamefont {X.~B.}\ \bibnamefont {Qiang}},
		\bibinfo {author} {\bibfnamefont {H.~Z.}\ \bibnamefont {Lu}}, \ and\ \bibinfo
		{author} {\bibfnamefont {X.~C.}\ \bibnamefont {Xie}},\ }\bibfield  {title}
	{\enquote {\bibinfo {title} {Coulomb instabilities of a three-dimensional
				higher-order topological insulator},}\ }\href {\doibase
		10.1103/PhysRevLett.127.176601} {\bibfield  {journal} {\bibinfo  {journal}
			{Phys. Rev. Lett.}\ }\textbf {\bibinfo {volume} {127}},\ \bibinfo {pages}
		{176601} (\bibinfo {year} {2021})}\BibitemShut {NoStop}%
	\bibitem [{\citenamefont {Araki}\ \emph {et~al.}(2020)\citenamefont {Araki},
		\citenamefont {Mizoguchi},\ and\ \citenamefont
		{Hatsugai}}]{PhysRevResearch.2.012009}%
	\BibitemOpen
	\bibfield  {author} {\bibinfo {author} {\bibfnamefont {H.}~\bibnamefont
			{Araki}}, \bibinfo {author} {\bibfnamefont {T.}~\bibnamefont {Mizoguchi}}, \
		and\ \bibinfo {author} {\bibfnamefont {Y.}~\bibnamefont {Hatsugai}},\
	}\bibfield  {title} {\enquote {\bibinfo {title} {$z_{Q}$ {B}erry phase for
				higher-order symmetry-protected topological phases},}\ }\href {\doibase
		10.1103/PhysRevResearch.2.012009} {\bibfield  {journal} {\bibinfo  {journal}
			{Phys. Rev. Res.}\ }\textbf {\bibinfo {volume} {2}},\ \bibinfo {pages}
		{012009} (\bibinfo {year} {2020})}\BibitemShut {NoStop}%
	\bibitem [{\citenamefont {Bibo}\ \emph {et~al.}(2020)\citenamefont {Bibo},
		\citenamefont {Lovas}, \citenamefont {You}, \citenamefont {Grusdt},\ and\
		\citenamefont {Pollmann}}]{PhysRevB.102.041126}%
	\BibitemOpen
	\bibfield  {author} {\bibinfo {author} {\bibfnamefont {J.}~\bibnamefont
			{Bibo}}, \bibinfo {author} {\bibfnamefont {I.}~\bibnamefont {Lovas}},
		\bibinfo {author} {\bibfnamefont {Y.}~\bibnamefont {You}}, \bibinfo {author}
		{\bibfnamefont {F.}~\bibnamefont {Grusdt}}, \ and\ \bibinfo {author}
		{\bibfnamefont {F.}~\bibnamefont {Pollmann}},\ }\bibfield  {title} {\enquote
		{\bibinfo {title} {Fractional corner charges in a two-dimensional
				superlattice {B}ose-{H}ubbard model},}\ }\href {\doibase
		10.1103/PhysRevB.102.041126} {\bibfield  {journal} {\bibinfo  {journal}
			{Phys. Rev. B}\ }\textbf {\bibinfo {volume} {102}},\ \bibinfo {pages}
		{041126} (\bibinfo {year} {2020})}\BibitemShut {NoStop}%
	\bibitem [{\citenamefont {Fraxanet}\ \emph {et~al.}(2023)\citenamefont
		{Fraxanet}, \citenamefont {Dauphin}, \citenamefont {Lewenstein},
		\citenamefont {Barbiero},\ and\ \citenamefont
		{Gonz\'alez-Cuadra}}]{PhysRevLett.131.263001}%
	\BibitemOpen
	\bibfield  {author} {\bibinfo {author} {\bibfnamefont {J.}~\bibnamefont
			{Fraxanet}}, \bibinfo {author} {\bibfnamefont {A.}~\bibnamefont {Dauphin}},
		\bibinfo {author} {\bibfnamefont {M.}~\bibnamefont {Lewenstein}}, \bibinfo
		{author} {\bibfnamefont {L.}~\bibnamefont {Barbiero}}, \ and\ \bibinfo
		{author} {\bibfnamefont {D.}~\bibnamefont {Gonz\'alez-Cuadra}},\ }\bibfield
	{title} {\enquote {\bibinfo {title} {Higher-order topological peierls
				insulator in a two-dimensional atom-cavity system},}\ }\href {\doibase
		10.1103/PhysRevLett.131.263001} {\bibfield  {journal} {\bibinfo  {journal}
			{Phys. Rev. Lett.}\ }\textbf {\bibinfo {volume} {131}},\ \bibinfo {pages}
		{263001} (\bibinfo {year} {2023})}\BibitemShut {NoStop}%
	\bibitem [{\citenamefont {Winkler}\ \emph {et~al.}(2006)\citenamefont
		{Winkler}, \citenamefont {Thalhammer}, \citenamefont {Lang}, \citenamefont
		{Grimm}, \citenamefont {Hecker~Denschlag}, \citenamefont {Daley},
		\citenamefont {Kantian}, \citenamefont {B\"{u}chler},\ and\ \citenamefont
		{Zoller}}]{winkler2006repulsively}%
	\BibitemOpen
	\bibfield  {author} {\bibinfo {author} {\bibfnamefont {K.}~\bibnamefont
			{Winkler}}, \bibinfo {author} {\bibfnamefont {G.}~\bibnamefont {Thalhammer}},
		\bibinfo {author} {\bibfnamefont {F.}~\bibnamefont {Lang}}, \bibinfo {author}
		{\bibfnamefont {R.}~\bibnamefont {Grimm}}, \bibinfo {author} {\bibfnamefont
			{J.}~\bibnamefont {Hecker~Denschlag}}, \bibinfo {author} {\bibfnamefont
			{A.~J.}\ \bibnamefont {Daley}}, \bibinfo {author} {\bibfnamefont
			{A.}~\bibnamefont {Kantian}}, \bibinfo {author} {\bibfnamefont {H.~P.}\
			\bibnamefont {B\"{u}chler}}, \ and\ \bibinfo {author} {\bibfnamefont
			{P.}~\bibnamefont {Zoller}},\ }\bibfield  {title} {\enquote {\bibinfo {title}
			{Repulsively bound atom pairs in an optical lattice},}\ }\href
	{http://dx.doi.org/10.1038/nature04918} {\bibfield  {journal} {\bibinfo
			{journal} {Nature}\ }\textbf {\bibinfo {volume} {441}},\ \bibinfo {pages}
		{853} (\bibinfo {year} {2006})}\BibitemShut {NoStop}%
	\bibitem [{\citenamefont {Salerno}\ \emph {et~al.}(2018)\citenamefont
		{Salerno}, \citenamefont {Di~Liberto}, \citenamefont {Menotti},\ and\
		\citenamefont {Carusotto}}]{salerno2018topological}%
	\BibitemOpen
	\bibfield  {author} {\bibinfo {author} {\bibfnamefont {G.}~\bibnamefont
			{Salerno}}, \bibinfo {author} {\bibfnamefont {M.}~\bibnamefont {Di~Liberto}},
		\bibinfo {author} {\bibfnamefont {C.}~\bibnamefont {Menotti}}, \ and\
		\bibinfo {author} {\bibfnamefont {I.}~\bibnamefont {Carusotto}},\ }\bibfield
	{title} {\enquote {\bibinfo {title} {Topological two-body bound states in the
				interacting {H}aldane model},}\ }\href
	{http://dx.doi.org/10.1103/PhysRevA.97.013637} {\bibfield  {journal}
		{\bibinfo  {journal} {Phys. Rev. A}\ }\textbf {\bibinfo {volume} {97}},\
		\bibinfo {pages} {013637} (\bibinfo {year} {2018})}\BibitemShut {NoStop}%
	\bibitem [{\citenamefont {Lyubarov}\ and\ \citenamefont
		{Poddubny}(2019)}]{lyubarov2019edge}%
	\BibitemOpen
	\bibfield  {author} {\bibinfo {author} {\bibfnamefont {M.}~\bibnamefont
			{Lyubarov}}\ and\ \bibinfo {author} {\bibfnamefont {A.}~\bibnamefont
			{Poddubny}},\ }\bibfield  {title} {\enquote {\bibinfo {title} {Edge states of
				photon pairs in cavity arrays with spatially modulated nonlinearity},}\
	}\href {http://dx.doi.org/10.1103/PhysRevA.100.053813} {\bibfield  {journal}
		{\bibinfo  {journal} {Phys. Rev. A}\ }\textbf {\bibinfo {volume} {100}}
		(\bibinfo {year} {2019})}\BibitemShut {NoStop}%
	\bibitem [{\citenamefont {Lin}\ \emph {et~al.}(2020)\citenamefont {Lin},
		\citenamefont {Ke},\ and\ \citenamefont {Lee}}]{lin2020interaction}%
	\BibitemOpen
	\bibfield  {author} {\bibinfo {author} {\bibfnamefont {L.}~\bibnamefont
			{Lin}}, \bibinfo {author} {\bibfnamefont {Y.}~\bibnamefont {Ke}}, \ and\
		\bibinfo {author} {\bibfnamefont {C.}~\bibnamefont {Lee}},\ }\bibfield
	{title} {\enquote {\bibinfo {title} {Interaction-induced topological bound
				states and {T}houless pumping in a one-dimensional optical lattice},}\ }\href
	{http://dx.doi.org/10.1103/PhysRevA.101.023620} {\bibfield  {journal}
		{\bibinfo  {journal} {Physical Review A}\ }\textbf {\bibinfo {volume} {101}}
		(\bibinfo {year} {2020})}\BibitemShut {NoStop}%
	\bibitem [{\citenamefont {Stepanenko}\ and\ \citenamefont
		{Gorlach}(2020)}]{stepanenko2020interaction}%
	\BibitemOpen
	\bibfield  {author} {\bibinfo {author} {\bibfnamefont {A.~A.}\ \bibnamefont
			{Stepanenko}}\ and\ \bibinfo {author} {\bibfnamefont {M.~A.}\ \bibnamefont
			{Gorlach}},\ }\bibfield  {title} {\enquote {\bibinfo {title}
			{Interaction-induced topological states of photon pairs},}\ }\href {\doibase
		10.1103/PhysRevA.102.013510} {\bibfield  {journal} {\bibinfo  {journal}
			{Phys. Rev. A}\ }\textbf {\bibinfo {volume} {102}},\ \bibinfo {pages}
		{013510} (\bibinfo {year} {2020})}\BibitemShut {NoStop}%
	\bibitem [{\citenamefont {Gorlach}\ and\ \citenamefont
		{Poddubny}(2017)}]{gorlach2017topological}%
	\BibitemOpen
	\bibfield  {author} {\bibinfo {author} {\bibfnamefont {M.~A.}\ \bibnamefont
			{Gorlach}}\ and\ \bibinfo {author} {\bibfnamefont {Al.~N.}\ \bibnamefont
			{Poddubny}},\ }\bibfield  {title} {\enquote {\bibinfo {title} {Topological
				edge states of bound photon pairs},}\ }\href
	{http://dx.doi.org/10.1103/PhysRevA.95.053866} {\bibfield  {journal}
		{\bibinfo  {journal} {Phys. Rev. A}\ }\textbf {\bibinfo {volume} {95}},\
		\bibinfo {pages} {053866} (\bibinfo {year} {2017})}\BibitemShut {NoStop}%
	\bibitem [{\citenamefont {Olekhno}\ \emph {et~al.}(2020)\citenamefont
		{Olekhno}, \citenamefont {Kretov}, \citenamefont {Stepanenko}, \citenamefont
		{Ivanova}, \citenamefont {Yaroshenko}, \citenamefont {Puhtina}, \citenamefont
		{Filonov}, \citenamefont {Cappello}, \citenamefont {Matekovits},\ and\
		\citenamefont {Gorlach}}]{olekhno2020topological}%
	\BibitemOpen
	\bibfield  {author} {\bibinfo {author} {\bibfnamefont {N.~A.}\ \bibnamefont
			{Olekhno}}, \bibinfo {author} {\bibfnamefont {E.~I.}\ \bibnamefont {Kretov}},
		\bibinfo {author} {\bibfnamefont {A.~A.}\ \bibnamefont {Stepanenko}},
		\bibinfo {author} {\bibfnamefont {P.~A.}\ \bibnamefont {Ivanova}}, \bibinfo
		{author} {\bibfnamefont {V.~V.}\ \bibnamefont {Yaroshenko}}, \bibinfo
		{author} {\bibfnamefont {E.~M.}\ \bibnamefont {Puhtina}}, \bibinfo {author}
		{\bibfnamefont {D.~S.}\ \bibnamefont {Filonov}}, \bibinfo {author}
		{\bibfnamefont {B.}~\bibnamefont {Cappello}}, \bibinfo {author}
		{\bibfnamefont {L.}~\bibnamefont {Matekovits}}, \ and\ \bibinfo {author}
		{\bibfnamefont {M.~A.}\ \bibnamefont {Gorlach}},\ }\bibfield  {title}
	{\enquote {\bibinfo {title} {Topological edge states of interacting photon
				pairs emulated in a topolectrical circuit},}\ }\href
	{http://dx.doi.org/10.1038/s41467-020-14994-7} {\bibfield  {journal}
		{\bibinfo  {journal} {Nat. Commun.}\ }\textbf {\bibinfo {volume} {11}},\
		\bibinfo {pages} {1436} (\bibinfo {year} {2020})}\BibitemShut {NoStop}%
	\bibitem [{\citenamefont {Stepanenko}\ \emph {et~al.}(2020)\citenamefont
		{Stepanenko}, \citenamefont {Lyubarov},\ and\ \citenamefont
		{Gorlach}}]{stepanenko2020topological}%
	\BibitemOpen
	\bibfield  {author} {\bibinfo {author} {\bibfnamefont {A.~A.}\ \bibnamefont
			{Stepanenko}}, \bibinfo {author} {\bibfnamefont {M.~D.}\ \bibnamefont
			{Lyubarov}}, \ and\ \bibinfo {author} {\bibfnamefont {M.~A.}\ \bibnamefont
			{Gorlach}},\ }\bibfield  {title} {\enquote {\bibinfo {title} {Topological
				states in qubit arrays induced by density-dependent coupling},}\ }\href
	{http://dx.doi.org/10.1103/PhysRevApplied.14.064040} {\bibfield  {journal}
		{\bibinfo  {journal} {Phys. Rev. Appl.}\ }\textbf {\bibinfo {volume} {14}},\
		\bibinfo {pages} {064040} (\bibinfo {year} {2020})}\BibitemShut {NoStop}%
	\bibitem [{\citenamefont {Stepanenko}\ \emph {et~al.}(2022)\citenamefont
		{Stepanenko}, \citenamefont {Lyubarov},\ and\ \citenamefont
		{Gorlach}}]{PhysRevLett.128.213903}%
	\BibitemOpen
	\bibfield  {author} {\bibinfo {author} {\bibfnamefont {A.~A.}\ \bibnamefont
			{Stepanenko}}, \bibinfo {author} {\bibfnamefont {M.~D.}\ \bibnamefont
			{Lyubarov}}, \ and\ \bibinfo {author} {\bibfnamefont {M.~A.}\ \bibnamefont
			{Gorlach}},\ }\bibfield  {title} {\enquote {\bibinfo {title} {Higher-order
				topological phase of interacting photon pairs},}\ }\href {\doibase
		10.1103/PhysRevLett.128.213903} {\bibfield  {journal} {\bibinfo  {journal}
			{Phys. Rev. Lett.}\ }\textbf {\bibinfo {volume} {128}},\ \bibinfo {pages}
		{213903} (\bibinfo {year} {2022})}\BibitemShut {NoStop}%
	\bibitem [{\citenamefont {Bukov}\ \emph {et~al.}(2015)\citenamefont {Bukov},
		\citenamefont {D{\textquotesingle}Alessio},\ and\ \citenamefont
		{Polkovnikov}}]{Bukov2015}%
	\BibitemOpen
	\bibfield  {author} {\bibinfo {author} {\bibfnamefont {M.}~\bibnamefont
			{Bukov}}, \bibinfo {author} {\bibfnamefont {L.}~\bibnamefont
			{D{\textquotesingle}Alessio}}, \ and\ \bibinfo {author} {\bibfnamefont
			{A.}~\bibnamefont {Polkovnikov}},\ }\bibfield  {title} {\enquote {\bibinfo
			{title} {Universal high-frequency behavior of periodically driven systems:
				{F}rom dynamical stabilization to {F}loquet engineering},}\ }\href {\doibase
		10.1080/00018732.2015.1055918} {\bibfield  {journal} {\bibinfo  {journal}
			{Adv. Phys.}\ }\textbf {\bibinfo {volume} {64}},\ \bibinfo {pages} {139}
		(\bibinfo {year} {2015})}\BibitemShut {NoStop}%
	\bibitem [{\citenamefont {Eckardt}\ and\ \citenamefont
		{Anisimovas}(2015)}]{Eckardt2015}%
	\BibitemOpen
	\bibfield  {author} {\bibinfo {author} {\bibfnamefont {A.}~\bibnamefont
			{Eckardt}}\ and\ \bibinfo {author} {\bibfnamefont {E.}~\bibnamefont
			{Anisimovas}},\ }\bibfield  {title} {\enquote {\bibinfo {title}
			{High-frequency approximation for periodically driven quantum systems from a
				{F}loquet-space perspective},}\ }\href {\doibase
		10.1088/1367-2630/17/9/093039} {\bibfield  {journal} {\bibinfo  {journal}
			{New J. Phys.}\ }\textbf {\bibinfo {volume} {17}},\ \bibinfo {pages} {093039}
		(\bibinfo {year} {2015})}\BibitemShut {NoStop}%
	\bibitem [{\citenamefont {Mikami}\ \emph {et~al.}(2016)\citenamefont {Mikami},
		\citenamefont {Kitamura}, \citenamefont {Yasuda}, \citenamefont {Tsuji},
		\citenamefont {Oka},\ and\ \citenamefont {Aoki}}]{PhysRevB.93.144307}%
	\BibitemOpen
	\bibfield  {author} {\bibinfo {author} {\bibfnamefont {T.}~\bibnamefont
			{Mikami}}, \bibinfo {author} {\bibfnamefont {S.}~\bibnamefont {Kitamura}},
		\bibinfo {author} {\bibfnamefont {K.}~\bibnamefont {Yasuda}}, \bibinfo
		{author} {\bibfnamefont {N.}~\bibnamefont {Tsuji}}, \bibinfo {author}
		{\bibfnamefont {T.}~\bibnamefont {Oka}}, \ and\ \bibinfo {author}
		{\bibfnamefont {H.}~\bibnamefont {Aoki}},\ }\bibfield  {title} {\enquote
		{\bibinfo {title} {Brillouin-{W}igner theory for high-frequency expansion in
				periodically driven systems: {A}pplication to {F}loquet topological
				insulators},}\ }\href {\doibase 10.1103/PhysRevB.93.144307} {\bibfield
		{journal} {\bibinfo  {journal} {Phys. Rev. B}\ }\textbf {\bibinfo {volume}
			{93}},\ \bibinfo {pages} {144307} (\bibinfo {year} {2016})}\BibitemShut
	{NoStop}%
	\bibitem [{\citenamefont {Mori}(2018)}]{PhysRevB.98.104303}%
	\BibitemOpen
	\bibfield  {author} {\bibinfo {author} {\bibfnamefont {T.}~\bibnamefont
			{Mori}},\ }\bibfield  {title} {\enquote {\bibinfo {title} {Floquet
				prethermalization in periodically driven classical spin systems},}\ }\href
	{\doibase 10.1103/PhysRevB.98.104303} {\bibfield  {journal} {\bibinfo
			{journal} {Phys. Rev. B}\ }\textbf {\bibinfo {volume} {98}},\ \bibinfo
		{pages} {104303} (\bibinfo {year} {2018})}\BibitemShut {NoStop}%
	\bibitem [{\citenamefont {Claeys}\ \emph {et~al.}(2018)\citenamefont {Claeys},
		\citenamefont {De~Baerdemacker}, \citenamefont {Araby},\ and\ \citenamefont
		{Caux}}]{PhysRevLett.121.080401}%
	\BibitemOpen
	\bibfield  {author} {\bibinfo {author} {\bibfnamefont {P.~W.}\ \bibnamefont
			{Claeys}}, \bibinfo {author} {\bibfnamefont {S.}~\bibnamefont
			{De~Baerdemacker}}, \bibinfo {author} {\bibfnamefont {O.~E.}\ \bibnamefont
			{Araby}}, \ and\ \bibinfo {author} {\bibfnamefont {J.-S.}\ \bibnamefont
			{Caux}},\ }\bibfield  {title} {\enquote {\bibinfo {title} {Spin polarization
				through {F}loquet resonances in a driven central spin model},}\ }\href
	{\doibase 10.1103/PhysRevLett.121.080401} {\bibfield  {journal} {\bibinfo
			{journal} {Phys. Rev. Lett.}\ }\textbf {\bibinfo {volume} {121}},\ \bibinfo
		{pages} {080401} (\bibinfo {year} {2018})}\BibitemShut {NoStop}%
	\bibitem [{\citenamefont {Eckardt}(2017)}]{RevModPhys.89.011004}%
	\BibitemOpen
	\bibfield  {author} {\bibinfo {author} {\bibfnamefont {Andr\'e}\ \bibnamefont
			{Eckardt}},\ }\bibfield  {title} {\enquote {\bibinfo {title} {Colloquium:
				{A}tomic quantum gases in periodically driven optical lattices},}\ }\href
	{\doibase 10.1103/RevModPhys.89.011004} {\bibfield  {journal} {\bibinfo
			{journal} {Rev. Mod. Phys.}\ }\textbf {\bibinfo {volume} {89}},\ \bibinfo
		{pages} {011004} (\bibinfo {year} {2017})}\BibitemShut {NoStop}%
	\bibitem [{\citenamefont {Jin}\ and\ \citenamefont
		{Knolle}(2024)}]{PhysRevResearch.6.L042033}%
	\BibitemOpen
	\bibfield  {author} {\bibinfo {author} {\bibfnamefont {H.-K.}\ \bibnamefont
			{Jin}}\ and\ \bibinfo {author} {\bibfnamefont {J.}~\bibnamefont {Knolle}},\
	}\bibfield  {title} {\enquote {\bibinfo {title} {Floquet prethermal order by
				disorder},}\ }\href {\doibase 10.1103/PhysRevResearch.6.L042033} {\bibfield
		{journal} {\bibinfo  {journal} {Phys. Rev. Res.}\ }\textbf {\bibinfo {volume}
			{6}},\ \bibinfo {pages} {L042033} (\bibinfo {year} {2024})}\BibitemShut
	{NoStop}%
	\bibitem [{\citenamefont {Bir}\ and\ \citenamefont {Pikus}(1974)}]{Bir1974}%
	\BibitemOpen
	\bibfield  {author} {\bibinfo {author} {\bibfnamefont {G.~L.}\ \bibnamefont
			{Bir}}\ and\ \bibinfo {author} {\bibfnamefont {G.}~\bibnamefont {Pikus}},\
	}\href@noop {} {\emph {\bibinfo {title} {Summetry and Strain-Induced Effects
				in Semiconductors}}}\ (\bibinfo  {publisher} {Keter, Jerusalem},\ \bibinfo
	{year} {1974})\BibitemShut {NoStop}%
	\bibitem [{\citenamefont {Cohen-Tannoudji}\ \emph {et~al.}(1998)\citenamefont
		{Cohen-Tannoudji}, \citenamefont {Dupont-Roc},\ and\ \citenamefont
		{Grynberg}}]{CCohenTannoudji1Atom}%
	\BibitemOpen
	\bibfield  {author} {\bibinfo {author} {\bibfnamefont {C.}~\bibnamefont
			{Cohen-Tannoudji}}, \bibinfo {author} {\bibfnamefont {J.}~\bibnamefont
			{Dupont-Roc}}, \ and\ \bibinfo {author} {\bibfnamefont {G.}~\bibnamefont
			{Grynberg}},\ }\href@noop {} {\emph {\bibinfo {title} {Atom-Photon
				Interactions}}}\ (\bibinfo  {publisher} {John Wiley and Sons},\ \bibinfo
	{year} {1998})\BibitemShut {NoStop}%
	\bibitem [{\citenamefont {Tamm}(1932)}]{Shockley1932}%
	\BibitemOpen
	\bibfield  {author} {\bibinfo {author} {\bibfnamefont {I.}~\bibnamefont
			{Tamm}},\ }\href@noop {} {\bibfield  {journal} {\bibinfo  {journal} {Phys. Z.
				Soviet Union.}\ }\textbf {\bibinfo {volume} {1}},\ \bibinfo {pages} {733}
		(\bibinfo {year} {1932})}\BibitemShut {NoStop}%
	\bibitem [{\citenamefont {Shockley}(1939)}]{PhysRev.56.317}%
	\BibitemOpen
	\bibfield  {author} {\bibinfo {author} {\bibfnamefont {W.}~\bibnamefont
			{Shockley}},\ }\bibfield  {title} {\enquote {\bibinfo {title} {On the surface
				states associated with a periodic potential},}\ }\href {\doibase
		10.1103/PhysRev.56.317} {\bibfield  {journal} {\bibinfo  {journal} {Phys.
				Rev.}\ }\textbf {\bibinfo {volume} {56}},\ \bibinfo {pages} {317} (\bibinfo
		{year} {1939})}\BibitemShut {NoStop}%
	\bibitem [{\citenamefont {Roushan}\ \emph {et~al.}(2016)\citenamefont
		{Roushan}, \citenamefont {Neill}, \citenamefont {Megrant}, \citenamefont
		{Chen}, \citenamefont {Babbush}, \citenamefont {Barends}, \citenamefont
		{Campbell}, \citenamefont {Chen}, \citenamefont {Chiaro}, \citenamefont
		{Dunsworth}, \citenamefont {Fowler}, \citenamefont {Jeffrey}, \citenamefont
		{Kelly}, \citenamefont {Lucero}, \citenamefont {Mutus}, \citenamefont
		{O’Malley}, \citenamefont {Neeley}, \citenamefont {Quintana}, \citenamefont
		{Sank}, \citenamefont {Vainsencher}, \citenamefont {Wenner}, \citenamefont
		{White}, \citenamefont {Kapit}, \citenamefont {Neven},\ and\ \citenamefont
		{Martinis}}]{Roushan2016}%
	\BibitemOpen
	\bibfield  {author} {\bibinfo {author} {\bibfnamefont {P.}~\bibnamefont
			{Roushan}}, \bibinfo {author} {\bibfnamefont {C.}~\bibnamefont {Neill}},
		\bibinfo {author} {\bibfnamefont {A.}~\bibnamefont {Megrant}}, \bibinfo
		{author} {\bibfnamefont {Y.}~\bibnamefont {Chen}}, \bibinfo {author}
		{\bibfnamefont {R.}~\bibnamefont {Babbush}}, \bibinfo {author} {\bibfnamefont
			{R.}~\bibnamefont {Barends}}, \bibinfo {author} {\bibfnamefont
			{B.}~\bibnamefont {Campbell}}, \bibinfo {author} {\bibfnamefont
			{Z.}~\bibnamefont {Chen}}, \bibinfo {author} {\bibfnamefont {B.}~\bibnamefont
			{Chiaro}}, \bibinfo {author} {\bibfnamefont {A.}~\bibnamefont {Dunsworth}},
		\bibinfo {author} {\bibfnamefont {A.}~\bibnamefont {Fowler}}, \bibinfo
		{author} {\bibfnamefont {E.}~\bibnamefont {Jeffrey}}, \bibinfo {author}
		{\bibfnamefont {J.}~\bibnamefont {Kelly}}, \bibinfo {author} {\bibfnamefont
			{E.}~\bibnamefont {Lucero}}, \bibinfo {author} {\bibfnamefont
			{J.}~\bibnamefont {Mutus}}, \bibinfo {author} {\bibfnamefont {P.~J.~J.}\
			\bibnamefont {O’Malley}}, \bibinfo {author} {\bibfnamefont
			{M.}~\bibnamefont {Neeley}}, \bibinfo {author} {\bibfnamefont
			{C.}~\bibnamefont {Quintana}}, \bibinfo {author} {\bibfnamefont
			{D.}~\bibnamefont {Sank}}, \bibinfo {author} {\bibfnamefont {A.}~\bibnamefont
			{Vainsencher}}, \bibinfo {author} {\bibfnamefont {J.}~\bibnamefont {Wenner}},
		\bibinfo {author} {\bibfnamefont {T.}~\bibnamefont {White}}, \bibinfo
		{author} {\bibfnamefont {E.}~\bibnamefont {Kapit}}, \bibinfo {author}
		{\bibfnamefont {H.}~\bibnamefont {Neven}}, \ and\ \bibinfo {author}
		{\bibfnamefont {J.}~\bibnamefont {Martinis}},\ }\bibfield  {title} {\enquote
		{\bibinfo {title} {Chiral ground-state currents of interacting photons in a
				synthetic magnetic field},}\ }\href {\doibase 10.1038/nphys3930} {\bibfield
		{journal} {\bibinfo  {journal} {Nat. Phys.}\ }\textbf {\bibinfo {volume}
			{13}},\ \bibinfo {pages} {146} (\bibinfo {year} {2016})}\BibitemShut
	{NoStop}%
	\bibitem [{\citenamefont {J\"unemann}\ \emph {et~al.}(2017)\citenamefont
		{J\"unemann}, \citenamefont {Piga}, \citenamefont {Ran}, \citenamefont
		{Lewenstein}, \citenamefont {Rizzi},\ and\ \citenamefont
		{Bermudez}}]{PhysRevX.7.031057}%
	\BibitemOpen
	\bibfield  {author} {\bibinfo {author} {\bibfnamefont {J.}~\bibnamefont
			{J\"unemann}}, \bibinfo {author} {\bibfnamefont {A.}~\bibnamefont {Piga}},
		\bibinfo {author} {\bibfnamefont {S.-J.}\ \bibnamefont {Ran}}, \bibinfo
		{author} {\bibfnamefont {M.}~\bibnamefont {Lewenstein}}, \bibinfo {author}
		{\bibfnamefont {M.}~\bibnamefont {Rizzi}}, \ and\ \bibinfo {author}
		{\bibfnamefont {A.}~\bibnamefont {Bermudez}},\ }\bibfield  {title} {\enquote
		{\bibinfo {title} {Exploring interacting topological insulators with
				ultracold atoms: {T}he synthetic {C}reutz-{H}ubbard model},}\ }\href
	{\doibase 10.1103/PhysRevX.7.031057} {\bibfield  {journal} {\bibinfo
			{journal} {Phys. Rev. X}\ }\textbf {\bibinfo {volume} {7}},\ \bibinfo {pages}
		{031057} (\bibinfo {year} {2017})}\BibitemShut {NoStop}%
	\bibitem [{\citenamefont {Cooper}\ \emph {et~al.}(2019)\citenamefont {Cooper},
		\citenamefont {Dalibard},\ and\ \citenamefont
		{Spielman}}]{RevModPhys.91.015005}%
	\BibitemOpen
	\bibfield  {author} {\bibinfo {author} {\bibfnamefont {N.~R.}\ \bibnamefont
			{Cooper}}, \bibinfo {author} {\bibfnamefont {J.}~\bibnamefont {Dalibard}}, \
		and\ \bibinfo {author} {\bibfnamefont {I.~B.}\ \bibnamefont {Spielman}},\
	}\bibfield  {title} {\enquote {\bibinfo {title} {Topological bands for
				ultracold atoms},}\ }\href {\doibase 10.1103/RevModPhys.91.015005} {\bibfield
		{journal} {\bibinfo  {journal} {Rev. Mod. Phys.}\ }\textbf {\bibinfo
			{volume} {91}},\ \bibinfo {pages} {015005} (\bibinfo {year}
		{2019})}\BibitemShut {NoStop}%
	\bibitem [{\citenamefont {Koh}\ \emph {et~al.}(2022)\citenamefont {Koh},
		\citenamefont {Tai},\ and\ \citenamefont {Lee}}]{PhysRevLett.129.140502}%
	\BibitemOpen
	\bibfield  {author} {\bibinfo {author} {\bibfnamefont {J.~M.}\ \bibnamefont
			{Koh}}, \bibinfo {author} {\bibfnamefont {T.}~\bibnamefont {Tai}}, \ and\
		\bibinfo {author} {\bibfnamefont {C.~H.}\ \bibnamefont {Lee}},\ }\bibfield
	{title} {\enquote {\bibinfo {title} {Simulation of interaction-induced chiral
				topological dynamics on a digital quantum computer},}\ }\href {\doibase
		10.1103/PhysRevLett.129.140502} {\bibfield  {journal} {\bibinfo  {journal}
			{Phys. Rev. Lett.}\ }\textbf {\bibinfo {volume} {129}},\ \bibinfo {pages}
		{140502} (\bibinfo {year} {2022})}\BibitemShut {NoStop}%
\end{thebibliography}
%

\end{document}